# Beyond Code: The Multidimensional Impacts of Large Language Models in Software Development

Sardar Bonabi[1], Sarah Bana[2], Vijay Gurbaxani[1], Tingting Nian[1]

[1] The Paul Merage School of Business, University of California, Irvine (UCI)
Irvine, CA, U.S.A.

[2] The George L. Argyros School of Business & Economics, Chapman University
Orange, CA, U.S.A.

## Abstract

Large language models (LLMs) are poised to significantly impact software development, especially in the Open-Source Software (OSS) sector. To understand this impact, we first outline the mechanisms through which LLMs may influence OSS through code development, collaborative knowledge transfer, and skill development. We then empirically examine how LLMs affect OSS developers' work in these three key areas. Leveraging a natural experiment from a temporary ChatGPT ban in Italy, we employ a Difference-in-Differences framework with two-way fixed effects to analyze data from all OSS developers on GitHub in three similar countries—Italy, France, and Portugal—totaling 88,022 users. We find that access to ChatGPT increases developer productivity by 6.4%, knowledge sharing by 9.6%, and skill acquisition by 8.4%. These benefits vary significantly by user experience level: novice developers primarily experience productivity gains, whereas more experienced developers benefit more from improved knowledge sharing and accelerated skill acquisition. In addition, we find that LLM-assisted learning is highly context-dependent, with the greatest benefits observed in technically complex, fragmented, or rapidly evolving contexts. We show that the productivity effects of LLMs extend beyond direct code generation to include enhanced collaborative learning and knowledge exchange among developers—dynamics that are essential for gaining a holistic understanding of LLMs' impact in OSS. Our findings offer critical managerial implications: strategically deploying LLMs can accelerate novice developers' onboarding and productivity, empower intermediate developers to foster knowledge sharing and collaboration, and support rapid skill acquisition—together enhancing long-term organizational productivity and agility.

# 1 Introduction

Generative AI (GenAI) tools, particularly large language models (LLMs), are increasingly recognized for their potential to enhance white-collar productivity. Among various fields, software development stands out as an area where LLMs may have the most significant impact (Ma et al. 2023). The inherently digital nature of software development, well-defined syntax and structure, and fundamental reliance on automation collectively make it a highly promising field for LLM-driven solutions. While initial studies suggest that LLMs can significantly improve the efficiency of software developers in controlled experimental settings (Cui et al. 2024; Peng et al. 2023), there is a limited understanding of how these tools influence critical dimensions of real-world software development beyond just code development, including collaborative workflows, knowledge sharing, and the acquisition of new skills—all of which are vital to the success and quality of software projects.

To address this gap, we empirically examine how ChatGPT, the world's most popular LLM, affects real-world, large-scale outcomes for software developers. Our analysis focuses on Open-Source Software (OSS), an increasingly important knowledge asset in modern economies (Blind et al. 2021; Blind and Schubert 2024; Robbins et al. 2018) and a major driver of societal value (Hoffmann et al. 2024b). We focus on three key dimensions of impact: (1) effects on code development activities, (2) contributions to collaborative learning through knowledge sharing, and (3) the acquisition and use of new programming languages. Each of these dimensions holds distinct importance—code development propels the creation and refinement of software, knowledge sharing fosters collaboration and broader sharing of best practices, and skill acquisition enables developers to adapt and innovate. Crucially, these dimensions are interdependent: knowledge sharing and skill acquisition generate positive spillover effects that elevate long-term productivity at both the individual and organizational level. Enhanced knowledge sharing and accelerated skill acquisition can amplify long-term developer productivity by lowering onboarding barriers, deepening expertise, and strengthening team-level performance. Together, these three dimensions provide a more holistic, ecosystem-level understanding of how LLMs transform the software development landscape in practice.

To study each of these three dimensions, we analyze the activity of software developers on the world's largest OSS community, GitHub. Software developers on GitHub engage in collaborative software development primarily in two ways. First, by performing tasks that directly contribute code to projects—such as initiating repositories, committing code, and submitting pull requests. We use these activities as a proxy for productivity. Second, by reviewing and providing feedback on peers' contributions—through code reviews, issue discussions, and commenting on pull requests—which we use as a proxy for knowledge sharing. Moreover, participating in diverse projects and collaborative learning activities on GitHub helps users acquire new skills and programming languages. These measures of productivity, knowledge sharing, and skill acquisition provide a foundation for examining how LLMs like ChatGPT shape OSS developer outcomes.

Our identification strategy leverages a natural experiment stemming from the abrupt, four-week ban on ChatGPT in Italy—which was later lifted—to analyze the impacts of both the loss and subsequent restoration of LLM access among OSS developers. Unlike the gradual adoption of LLMs, this sudden, uniform loss of access acts as an exogenous shock, allowing us to isolate the effect of LLM access. Using a Difference-in-Differences (DiD) framework with two-way fixed effects, we compare the activity of all OSS developers in Italy (the treatment group) to that of developers in two similar countries, France and Portugal (the control groups), encompassing a total of 88,022 GitHub users. We also consider the possibility that developers may have switched to alternative LLMs or used VPNs, posing potential threats to the accuracy of our estimates. In Section 6 (Robustness Checks), we empirically demonstrate that such actions are unlikely to have materially affected our results.

Our findings provide significant new insights into how generative AI tools like ChatGPT reshape critical dimensions of OSS development—*code productivity*, *collaborative knowledge sharing*, and *skill acquisition*. We find that compared to the pre-treatment period, losing access to ChatGPT leads to a substantial 6.4% decrease in code development productivity, highlighting the magnitude of immediate efficiency gains offered by LLMs. We further identify an 8.4% decline in developers' skill acquisition, measured through their use of new programming languages, which is an aspect impacting productivity, often overlooked yet critical for sustained innovation and adaptability. Furthermore, our analysis shows that restoring ChatGPT access boosts knowledge sharing activities by 9.6%, capturing the important indirect effects LLMs have in facilitating collaboration, peer feedback, and community engagement. These findings underscore the multidimensional nature of productivity improvements driven by generative AI and illustrates how traditional productivity metrics alone fail to capture the comprehensive value LLMs deliver in software development.

Additionally, our results show that access to LLMs does not produce a uniform impact across all developers, presenting distinct implications for workforce management and talent development in the era of LLMs. We find productivity improvements from using LLMs are most pronounced for less experienced developers, narrowing the productivity gap between novice and experienced developers. For more experienced developers, however, ChatGPT's impacts are even more transformative, significantly enhancing their participation in knowledge sharing and accelerating their rate of skill acquisition. These results indicate that LLMs do more than simply automate routine tasks; they substantially reshape the learning dynamics and collaborative structure of OSS communities, generating long-term value that prior studies focused on productivity have not fully captured.

Our study extends existing research on the business value of AI in several ways. First, unlike prior studies that predominantly focus on short-term impacts of LLMs on productivity metrics (e.g., Kreitmeir and Raschky 2024; Peng et al. 2023), we show that LLMs have the potential to deliver substantial long-term benefits through indirect pathways such as accelerated learning and enhanced collaborative knowledge exchange. Second, while existing studies focus on the impacts of LLMs on knowledge sharing platforms such as Stack Overflow, StackExchange, and Reddit (e.g., Burtch et al. 2024; Quinn and Gutt 2023; Su et al. 2024) which primarily involve standalone question-answer interactions, our research extends the literature by showing how LLMs affect ongoing, more structured, and integrated collaborative work. Moreover, we contribute to the emerging literature on LLMs' impact on collaborative learning and skill acquisition by providing the first empirical evidence on how it shapes developers' skill acquisition. While existing research has raised concerns about potential negative effects such as superficial understanding or over-reliance (e.g., Bastani et al. 2024; Extance 2023), our findings demonstrate a strong positive impact of LLMs, particularly for mid-level developers acquiring new programming language skills in real-world OSS projects. Lastly, we extend the emerging literature on the context-dependent complementary role of LLM-assisted learning (Gao et al. 2023; Hemmer et al. 2024; Ye et al. 2023), by demonstrating that the benefits of LLMs for skill development are not evenly distributed across learning contexts. We show that LLMs provide the greatest support in contexts where developers face steep learning curves—namely, when engaging with technically complex, poorly documented, or rapidly evolving programming languages.

The remainder of the paper is structured as follows. In the next section, we lay the foundation for this study by reviewing the literature on LLMs within the software development domain. We then explore how these tools influence the activities and processes integral to modern collaborative software development. Building on these insights, we present a graphical representation of modern collaborative software development, which serves as the foundation for formulating and examining our research questions.

## 2 Literature Review

Our research contributes to three closely related streams of literature. First, we build on the growing body of work examining the business value of GenAI, particularly LLMs, in relation to worker productivity. Second, we contribute to the literature on the role of LLMs in knowledge sharing within online platforms. Third, we extend the emerging literature on LLMs' role in learning, particularly in supporting skill acquisition and collaborative learning in professional settings.

**2.1 Productivity Impacts of LLMs**

A few studies have explored the impact of adopting a specialized LLM (GitHub Copilot) on the productivity of software developers. In a controlled online experiment, Peng et al. (2023) find that developers with access to GitHub Copilot complete a programming task about 56% faster. Furthermore, a field experiment on two companies shows that software developers using GitHub Copilot complete about 26% more programming tasks (Cui et al. 2024). While LLMs are associated with overall productivity gains, these improvements are not uniform across different types of tasks. Software developers using GitHub Copilot experience greater benefits in routine tasks than those requiring more creative, non-standard problem solving (Yeverechyahu et al. 2024).

Aside from positive impact on the software developers' code development speed, LLMs also appear to influence the allocation of developers' efforts. Software developers with access to GitHub Copilot focus more on code development and less on project management tasks (Hoffmann et al. 2024a).

These studies have examined the impact of GitHub Copilot, a specialized coding LLM, primarily designed as a code-autocompletion tool embedded within integrated development environments (IDEs).[1] However, general-purpose LLMs such as ChatGPT differ fundamentally in their versatility and scope (Siroš et al. 2024). Unlike GitHub Copilot, ChatGPT supports interactive, conversational exchanges, enabling developers not only to generate code but also to receive guidance on project design and structure, ask conceptual questions, and debug (Barke et al. 2023). Thus, ChatGPT's impact on developer productivity, knowledge sharing, and skill acquisition may differ significantly from those observed with a coding-specific LLM such as GitHub Copilot.

While many studies explore the effects of the gradual *adoption* of LLMs, we study the impact of sudden *loss of access* to an LLM, a setting that we argue allows for more precise causal estimation. When users adopt a technology over time, their decision may correlate with unobserved factors like skill level, motivation, or project demands, making it difficult to isolate the true causal effect. Moreover, adoption occurs at different times across users. If the technology has a positive (or negative) effect, earlier adopters will experience it for longer, potentially biasing or diluting estimates when averaged across all users (Callaway and Sant'Anna 2021; de Chaisemartin and D'Haultfœuille 2020; Egami and Yamauchi 2023; Freedman et al. 2023). In contrast, a sudden, uniform loss of access functions as an exogenous *shock* that affects all users simultaneously, thereby minimizing self-selection and timing-related biases. This approach enables more precise estimation of the technology's causal effect.

---

[1] An integrated development platform (IDE) is a software tool that streamlines the coding process by combining editing, building, and debugging in one interface. At the time window of this study, GitHub Copilot was limited to code block auto completion. In December of 2023–several months after our study period–a conversational version of GitHub Copilot (GitHub Copilot Chat) was released.
https://github.blog/news-insights/product-news/github-copilot-chat-now-generally-available-for-organizations-and-individuals/

Perhaps the closest to our research, a study by Kreitmeir and Raschky (2024) examines the impact of an interruption in LLM access on software developers. They also analyze the effects of the ChatGPT ban in Italy by analyzing the activity of GitHub users to assess its impact on developer productivity. Unlike our study, which investigates more sustained and longer-term effects, their research focuses on *immediate outcomes* by comparing productivity metrics from four days before to four days after the ban. They find that losing access to ChatGPT has no significant overall impact on the quantity or quality of output from Italian users within their short window of study.[2] Moreover, they argue that upon losing access to ChatGPT, less experienced users see *an improvement* in output quantity and quality, suggesting that ChatGPT *adversely* affects lower-experience developers. However, this study does not address the longer-term productivity effects beyond the immediate aftermath of the ban. Because software development unfolds in iterative cycles—spanning design, coding, testing, and feedback—a brief observation period may overlook the broader impacts of LLM use. Consequently, we study developer activity over a longer timeframe, four months of activity—eight weeks before the ban, the four-week ban, and four weeks after—to capture more sustained and accurate effects that a short-term view may miss. Moreover, we consider not only productivity, but also the effects on knowledge sharing and byproducts such as skill acquisition, two critical components of collaborative open-source software development.

---

[2] While no significant overall impact was found on the quantity or quality of output following the ban, a notable negative effect was observed on the number of closed issues.

**2.2 Knowledge Sharing Impacts of LLMs**

To the best of our knowledge, no existing studies have examined the impact of generative AI on knowledge sharing within software development communities, particularly in the context of OSS development. In non-OSS domains, recent studies show that these tools impact how knowledge spreads across online communities. Burtch et al. (2024) show that the introduction of ChatGPT led to a decrease in the number of questions asked on Stack Overflow, a popular programming-focused Q&A platform, while having no effect on a general-purpose alternative, Reddit. Moreover, Quinn and Gutt (2023) show that the introduction of ChatGPT resulted in a decline in overall user engagement on an online knowledge-sharing platform, along with an increase in the complexity of posted questions. We study the effects of LLMs on knowledge sharing activities within the OSS community, where they remain understudied. Understanding these effects is particularly important because the transfer of knowledge may have spillover effects on the productivity of the developers (Chang and Gurbaxani 2012), and the OSS community's existence, advancement, and overall sustainability depend on developers' diverse knowledge contributions and collaborative learning.

**2.3 Learning Impacts of LLMs**

In addition to productivity and knowledge sharing, LLMs have the potential to influence users' learning and skill acquisition (Baidoo-Anu and Owusu Ansah 2023; Bastani et al. 2024; Chen et al. 2024; Chen et al. 2023; Chowdhury et al. 2024; Extance 2023; Lawasi et al. 2024; MacNeil et al. 2023; Nye et al. 2023). In the software development domain, the ability to write and modify code, understand its purpose, and articulate its functionality are fundamental skills that software developers must develop (Cunningham et al. 2022; Murphy et al. 2012; Wang et al. 2020). Chen et al. (2024) conducted a lab experiment showing that interacting with an LLM enhanced students' coding skill acquisition. However, these improvements were not uniform across all skill types; while overall coding skills improved, gains were notably limited in error correction and debugging tasks.

This positive impact on programming skill acquisition in students can be partly attributed to the LLMs' ability to explain concepts (MacNeil et al. 2023; MacNeil et al. 2022). Such explanations can help students understand blocks of code (Marwan et al. 2019). Moreover, explanations can improve the overall learning outcomes during the code debugging and correction process, thereby reducing the effort required to overcome these challenges (Griffin 2016). LLMs may play a particularly impactful role in shaping skill acquisition among software development learners because they can tailor explanations to each student's current position on the skill acquisition curve.

While preliminary studies have examined how LLMs influence students' skill acquisition in software development under controlled laboratory conditions, no research to date has explored their impact on working professionals, particularly in large-scale, real-world settings.

## 3 Research Context and Questions

We begin by identifying the key stages and activities in the OSS development process. Next, we examine how LLM integration may influence these activities. Building on this analysis, we then formulate the research questions that guide our study.

### 3.1 Research Context

The OSS development process can be conceptualized in three main stages, each characterized by distinct developer activities. First, Project Initiation begins when a developer creates a repository or joins an existing one, formally launching the project and establishing its collaborative workspace (*creating repository*). Second, Code Development encompasses the core engineering tasks of writing or modifying code (*code commit*) and submitting code for integration into the main project (*pull request*). Third, Community Engagement sustains and improves the project through peer review of pull requests (*pull request review*), reporting issues to flag bugs or suggest enhancements (*issue report*), and participating in threaded discussions to share knowledge and coordinate future work (*discussion*). Together, these six activities—repository creation, code commits, pull request, pull request review, issue report, and project discussion—not only drive the project's progress but also offer developers continuous opportunities to acquire new skills, such as learning new programming languages, frameworks, or architectural patterns (*skill acquisition*). Figure 1 presents a visual representation of modern collaborative OSS development.[3,4]

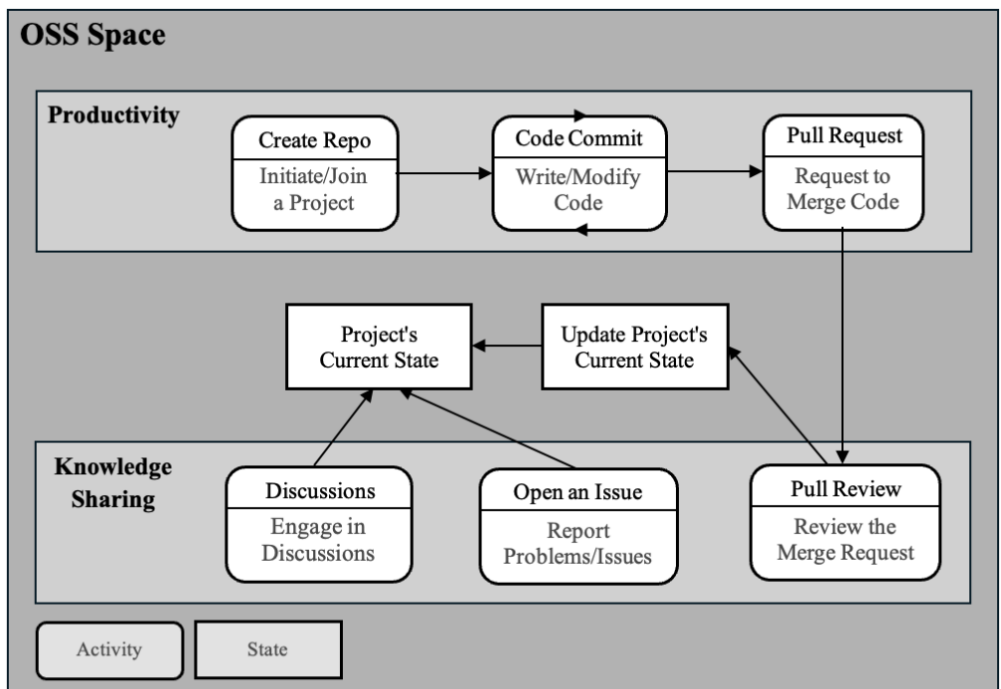

*Figure 1: Open-Source Software Development Process*

---

[3] We do not present this visual representation of modern collaborative software development as a contribution of our study. Rather, it is intended to help readers unfamiliar with OSS development understand its key activities and processes, and to establish a foundation for testing our hypotheses.

[4] On GitHub, users can also create *private repositories*, where access to the code and files is restricted to designated collaborators. For the purposes of this study, we focus exclusively on open-source and publicly available repositories.

We measure productivity using repository creation, code commits, and pull requests, as these activities directly contribute to the development and progression of OSS projects. Creating a repository establishes a new project space and framework for code storage and collaboration, an essential first step in software development. Committing code represents direct contributions to a project's development, with each commit reflecting incremental progress. Opening a pull request involves proposing changes to the project that, once reviewed, can be merged to advance the main codebase. This step is crucial for consistently integrating individual work efforts into the larger project. We aggregate these three activities into a composite variable, *Productivity*, as together, these activities are foundational to project advancement and closely associated with developer productivity.

Moreover, we measure knowledge sharing through three key activities; pull request reviews, discussions, and issue openings. These activities enable the exchange of insights, feedback, and solutions among project contributors. Reviewing pull requests involves assessing proposed code changes, providing explanations, and suggesting improvements, processes that help disseminate technical knowledge and best practices within the development team. Similarly, opening issues and initiating or participating in discussions provides a forum for discussing potential bugs, proposing features, and raising questions, thereby enabling community members to share expertise, learn from peers, and collaboratively address technical challenges. We aggregate these three activities into a composite variable, *Knowledge Sharing*, as these activities are central to enabling effective knowledge sharing among all participants, which is crucial in the project's continuous evolution.

Lastly, we construct the *Skill Acquisition* variable by tracking developers' use of new programming languages in their projects, which reflects their ongoing efforts to expand into new technical domains.

In the following subsections, we analyze how LLMs influence each of these three dimensions—productivity, knowledge sharing, and skill acquisition—and use these insights to formulate the research questions that guide our study.

## 3.2 Research Questions

In this study, examine the impact of LLMs three core dimensions of OSS development. Specifically, we address the following research questions:

**RQ1:** How do LLMs affect the *productivity* of software developers in the OSS community?

**RQ2:** How do LLMs impact the *knowledge sharing* among software developers in the OSS community?

**RQ3:** How do LLMs influence the *skill acquisition* of software developers in the OSS community?

These three research questions are central to our study because they encompass the core components of OSS development. Productivity determines the pace at which code is produced and integrated, directly affecting project advancement. Knowledge sharing sustains the collaborative model that distinguishes OSS, impacting code quality and collective problem-solving. Skill acquisition contributes to the community's long-term capacity to innovate by enabling contributors to adopt new programming languages, frameworks, and tools. LLMs have the potential to affect each of these dimensions—by automating routine coding tasks, streamlining peer review, and providing on-demand guidance—yet their overall impact remains unclear. Studying productivity, knowledge sharing, and skill acquisition together therefore enables a comprehensive assessment of LLMs' role in OSS and provides crucial insights relevant for researchers, developers, platform designers, team leads, technology managers, and policymakers seeking to understand how LLMs influence collaborative software development.

**3.2.1 Impact of LLMs on Productivity**

LLMs can impact each of the three components of developer productivity–repository creation, code commit, and pull request. LLMs can lower the barriers to entry for initiating new projects or joining existing ones by providing instant coding assistance, documentation generation, and problem-solving support. They help developers overcome initial challenges such as setting up project environments and understanding complex codebases. This reduction in initial complexity and increased accessibility may democratize participation in OSS and encourage more developers to launch new projects or contribute to ongoing ones,[5] thereby increasing overall engagement in OSS development.

Moreover, LLMs can impact developers' code writing activity–specifically, *code commit*– which is a key aspect of software developers' productivity (Vaithilingam et al. 2022). Developers can use these models to generate code (Ding et al. 2024), complete code segments, refactor existing code (Chavan et al. 2024), replace outdated functions with more efficient alternatives, address compiler errors (Bubeck et al. 2023), and create test cases (Ding et al. 2024; Haji et al. 2024). These capabilities may lead to more frequent and efficient code commits, as developers can implement changes more rapidly and with higher accuracy.

Lastly, LLMs can assist developers in preparing their code changes for integration into the main project by helping them create well-structured pull requests. This includes generating clear and comprehensive descriptions of proposed changes that supplement the pull request. Moreover,

---

[5] Using the methodology outlined in Sections 4 and 5, we examined the initiation activities of software developers. As shown in Table A1 in the Appendix, our analysis indicates that following the lifting of the ban, project initiation by software developers increased by 9.9%.

LLMs can assist the developers in writing documentation and inline comments aligned with the code, thereby enhancing the understandability and maintainability of the project's codebase (Ding et al. 2024). This function is particularly crucial in OSS development, where effective community engagement begins with understanding the current state of the project. Collectively, these enhancements may lead to improved developer productivity.

While the factors discussed above highlight ways LLMs may enhance developer productivity, several countervailing factors may reduce productivity—potentially outweighing the gains. First, interacting with an LLM introduces time and cognitive costs, as developers must interrupt their workflow to interact with the model, understand its responses, and integrate them into the codebase. Additionally, the non-deterministic nature of LLMs introduces the risk of suboptimal, incomplete, or incorrect responses (Baidoo-Anu and Owusu Ansah 2023; Bubeck et al. 2023; Vaithilingam et al. 2022)–a phenomenon commonly referred to as *LLM hallucination* (Bang et al. 2023). Managing this risk requires developers to spend additional time verifying the model's output before incorporating it into their projects, which can offset the initial productivity gains. Given these opposing effects on developer productivity, the net effect of LLMs on developer productivity remains unclear, and we aim to evaluate this impact through this study.

**3.2.2 Impact of LLMs on Knowledge Sharing**

LLMs can impact each of the three components of knowledge sharing–pull request review, discussion, and issue opening. LLMs can assist project maintainers during the pull review process by helping them better understand submitted code (Bubeck et al. 2023; Unterkalmsteiner et al. 2024). Specifically, during pull request reviews, LLMs can analyze submitted code for alignment with project objectives and standards, detect inefficiencies, and identify common programming errors. This support can streamline the review process and reduce the cognitive burden on human reviewers.

The use of LLMs can also impact the remaining two knowledge-sharing activities: participating in discussions and reporting issues. Understanding a project's codebase is a prerequisite for developers to participate in discussions or report issues, and is one of the most challenging aspects of collaborative software development (Dong et al. 2021). LLMs can help mitigate this challenge by providing detailed explanations that enable developers to navigate and comprehend complex code more easily (Bubeck et al. 2023). Moreover, LLMs can assist developers in articulating their thoughts more clearly and generating more precise and detailed issue reports and feature requests, potentially reducing the time required for these tasks. Furthermore, considering that English is the primary language of the OSS community, utilization of LLMs may lower entry barriers for non-English-speaking developers by providing translation support and helping them communicate more effectively in English. Collectively, these capabilities may improve developers' participation in knowledge-sharing activities.

While LLMs have the potential to enhance knowledge sharing within the OSS community, several factors may limit or even reverse this impact. First, as LLMs automate aspects of the code review process—such as error detection and improvement suggestion—developers may come to view their role in the review process as less essential. Given the voluntary nature of OSS contributions (Bitzer and Geishecker 2010), this perception shift could reduce their sense of responsibility, leading to decreased participation in reviews. The self-organizing structure of OSS development (Lindberg et al. 2024) further enables developers to redirect their efforts toward tasks they perceive as more engaging or impactful (Hoffmann et al. 2024a). As a result, reliance on LLMs for routine review tasks may lead to a decline in overall code review activity. In addition, developers may bypass collaborative processes of discussion and problem-solving altogether, opting instead to rely on LLM-generated solutions or automated issue detection. This change could also lead to a decrease in discussions and overall knowledge sharing around a project. Given these potential conflicting impacts, the overall impact of LLMs on knowledge sharing in OSS remains an open question—one that this study seeks to empirically investigate.

**3.2.3 Impact of LLMs on Skill Acquisition**

LLMs can also impact the skill acquisition of OSS developers. Developers have the opportunity to learn new programming languages through participating in OSS development. LLMs may enhance this learning process by explaining concepts and syntax in natural language (Bubeck et al. 2023), providing contextualized examples, answering questions, and offering real-time feedback on developers' code. Moreover, LLMs can personalize the learning experience by tailoring responses to the developer's current project context. Lastly, LLMs can also "translate" code from a familiar programming language into a new one (Korinek 2023), while explaining the syntactic differences and guiding necessary transitions. Ultimately, these capabilities may lead to improved skill acquisition among OSS developers.

However, LLMs may not always support skill acquisition and could, potentially, hinder it. LLMs may generate incorrect information (Bubeck et al. 2023; Vaithilingam et al. 2022), which can confuse or mislead learners. Moreover, in a controlled experiment, Bastani et al. (2024) show that learners who used an LLM during the learning phase performed significantly worse on subsequent examination phase—when the LLM was no longer available—than those in a control group. This effect may stem from overreliance on LLMs, which in turn may have an adverse impact on learning outcomes. Furthermore, the impersonal nature of LLM feedback lacks the contextual sensitivity and mentorship typically provided through human interactions (Luo 2024;

Yu and Guo 2023). Lastly, initial evidence also suggests that while AI may enhance learning for advanced users, it could hinder skill development for others (Toner-Rodgers 2024). Given these contrasting impacts of LLMs on skill acquisition, it remains unclear whether LLMs ultimately support or hinder software developers' learning, motivating the need for empirical investigation.

In the following sections, address the three research questions outlined in this study. The rest of the paper is organized as follows: Section 4 describes the data and empirical framework. Section 5 presents the main results. Section 6 details the robustness checks and their outcomes. Section 7 explores heterogeneity in effects across different subgroups of software developers. Section 8 analyzes the impact of LLMs on learning various programming languages and technologies. Finally, Section 9 concludes the study by summarizing the key findings and their implications.

# 4 Data and Empirical Framework

## 4.1 Data

To examine the impact of the ChatGPT ban in Italy and its subsequent lift on software developers within this country, we analyze the activity of software developers on the world's largest OSS platform, GitHub.[6] GitHub is widely used by developers to collaborate on projects, share code, and manage software development tasks, and has been extensively adopted in prior research (Kreitmeir and Raschky 2024; Malgonde et al. 2023; Peng et al. 2023; Yeverechyahu et al. 2024).

We collected activity data from GitHub users who listed their locations as Italy, France, or Portugal on their profiles covering a 16-week period from February 4, 2023, to May 26, 2023. This period includes eight weeks before the ban, the four-week ban period, and four weeks after the ban was lifted. The resulting dataset contains weekly activity data for 88,022 unique users. This dataset includes only public projects with open-source codebases, as the focus of the study is on the OSS community. In line with existing literature, we exclude users who showed no activity throughout the study's observation period, as they are likely inactive participants (Malgonde et al. 2023; Moqri et al. 2018).

---

[6] https://github.com/open-source, and https://en.wikipedia.org/wiki/GitHub.

For each user in the three countries, we collected three sets of data: (1) profile data, (2) activity data, and (3) programming language data. The **profile data** includes several attributes from user profiles, including username, location, company affiliation (indicator), number of public repositories,[7] number of public gists, number of followers, number of followings, profile creation date, and the date of last profile update.[8]

In addition, we collected **activity data** for the users in the panel, capturing six key types of OSS contributions: repository creations, code commits, pull requests, pull request reviews, issue reports, and discussion participations. Following the framework outlined in Section 3.1, we group these activities into two main constructs: (1) Productivity, which includes repository creations, code commits, and pull requests; and (2) Knowledge Sharing, which includes pull request reviews, issue reports, and discussions participations.

In addition to the profile and activity data, we collected all available information on the **programming languages** each user in the panel has ever used, from the date they joined the platform until the end of the observation window. This resulted in 1,989,535 repositories across 87,536 unique users.[9] Using this dataset, we measured the number of new programming languages a user employed in their projects during their tenure on the platform.[10] We constructed the dataset accordingly and refer to the resulting variable as *Skill Acquisition*. Table 1 provides a brief description of the key activities and the main variables in both the activity and skill acquisition datasets.

---

[7] A detailed explanation of GitHub-specific terms is provided in Table A2 of the Appendix.
[8] Summary statistics for these variables are presented in Table A3 of the Appendix.
[9] Among the 88,022 active users in the panel, 486 users did not have any public repositories.

**Table 1. Variable Descriptions**

| Variable | Description |
| --- | --- |
| Repository Creation | The number of new codebases initialized or established by the developer. |
| Code Commits | The number of code updates or changes submitted by the developer to an existing repository. |
| Pull Requests | The number of code change proposals submitted by the developer for integration into a project's codebase |
| Pull Request Reviews | The number of code change proposals evaluated or reviewed by the developer. |
| Issues Reports | The number of new entries made by the developer to report bugs, suggest enhancements, or raise questions. |
| Discussions | The number of conversation threads initiated or participated in by the developer to exchange ideas or coordinate project work. |
| **Productivity** | The sum of repositories created, commits made, and pull requests submitted by the developer. |
| **Knowledge Sharing** | The sum of pull requests reviewed, issues opened, and discussions participated in by the developer. |
| **Skill Acquisition** | The number of new programming languages used by the developer during the study period. A language is considered "new" if the developer had not used it in any prior project. |

Table 2 presents summary statistics for the activity and skill acquisition variables, first as individual components and then aggregated into productivity, knowledge sharing, and skill acquisition constructs. Among all activities, code commits are the most frequent, with developer in Italy averaging 5.33 commits per week, compared to 4.73 in France and Portugal. However, there is substantial variability across user-weeks, as indicated by large standard deviations (16.59 and 15.07, respectively), suggesting that some developers are significantly more active in some weeks than others. Other productivity-related activities–e.g., repository creation and pull requests–occur less frequently but follow similar patterns across the two groups.

---

[10] Additional details on the construction of the derived variable are provided in Section A1 of the Appendix.

Knowledge-sharing activities, including pull request reviews, issue openings, and discussions, are considerably less frequent. Discussions, in particular, are rare, averaging only 0.01 per week in both groups. When aggregated, productivity is slightly higher in Italy (5.92) than in France and Portugal (5.38), while knowledge-sharing levels are largely comparable. Skill acquisition, measured by the adoption of new programming languages, is also similar across groups (0.67 in Italy and 0.64 in France and Portugal). Overall, these statistics highlight the centrality of code commits in OSS activity, the high variability in user engagement across weeks, the relatively infrequent nature of knowledge-sharing activities, and the comparable rates of skill acquisition between groups.

**Table 2. Summary Statistics for the Activity and Skill Acquisition Datasets**

| Variables | Italy | | | France and Portugal | | |
|---|---|---|---|---|---|---|
| | Obs | Mean | SD | Obs | Mean | SD |
| **Productivity** | 133,400 | 5.92 | 17.10 | 446,895 | 5.38 | 15.79 |
| Repositories Creation | 133,400 | 0.25 | 0.71 | 446,895 | 0.22 | 0.70 |
| Code Commits | 133,400 | 5.33 | 16.59 | 446,895 | 4.73 | 15.07 |
| Pull Requests | 133,400 | 0.35 | 1.59 | 446,895 | 0.43 | 1.83 |
| **Knowledge Sharing** | 133,400 | 0.48 | 2.59 | 446,895 | 0.52 | 2.48 |
| Pull Reviews | 133,400 | 0.32 | 2.26 | 446,895 | 0.34 | 1.96 |
| Issue Reports | 133,400 | 0.16 | 1.00 | 446,895 | 0.18 | 1.26 |
| Discussions | 133,400 | 0.01 | 0.16 | 446,895 | 0.01 | 0.17 |
| **Skill Acquisition** | 429,926 | 0.67 | 1.95 | 1,559,609 | 0.64 | 1.93 |

In the next subsection, we present the empirical framework and identification strategy used to estimate the causal effects of the ChatGPT ban and its subsequent lift on the three main variables of this study: productivity, knowledge sharing, and skill acquisition.

**4.2 Empirical Framework**

To estimate the causal impact of access to ChatGPT on software development outcomes, we leverage a natural experiment arising from the temporary ban on ChatGPT in Italy. On March 31, 2023, the Italian government banned ChatGPT due to data privacy concerns, prompting OpenAI to disable access for users located in Italy (hereafter referred to as the "Ban").[11] Four weeks later, on April 28, 2023, the ban was lifted (hereafter referred to as the "Lift") after OpenAI addressed the government's concerns.[12]

To isolate the impacts of both the ban and the lift on developers in Italy from unobserved time-varying factors, we employ a Difference-in-Differences (DiD) approach, using users in France or Portugal as the control group. These three countries share similar geographic, economic, and regulatory characteristics, making their software development landscapes comparable. Geographically, all three countries are located in Europe. On the macroeconomic level, they have similar GDP per capita (see Figure A2 in the Appendix)[13] and a comparable share of the software industry within each country's GDP (ranging from 0.4% to 1% in 2021).[14] In addition, all three countries are members of the European Union and adhere to common regulations, such as the

---

[11] https://www.nytimes.com/2023/03/31/technology/chatgpt-italy-ban.html. The message displayed to users in Italy is presented in Figure A1 of the Appendix.
[12] https://www.bbc.com/news/technology-65431914
[13] https://data.worldbank.org/indicator/NY.GDP.PCAP.CD?contextual=default&end=2022&locations=IT-FR-PT&start=2000&view=chart
[14] https://www.ibisworld.com/france/industry-statistics/software-development/3595/

General Data Protection Regulation (GDPR)[15] and the Digital Services Act (DSA).[16] Furthermore, although English is the predominant language used on GitHub, none of these countries have English as a primary language, making them further comparable in terms of language-based barriers. Lastly, as we show in Section 5.4, these countries showed parallel trends in OSS development activity during the pre-treatment period, supporting the suitability of France and Portugal as the control group.

We compare the weekly activity outcomes of users in Italy, identified based on their self-reported location on GitHub, with those in the control group (France and Portugal) across three time periods: before the ban, during the ban, and after the ban was lifted. To estimate the average treatment effect on the treated developers (ATT) for our three main dependent variables, we use a Poisson regression model with two-way fixed effects. The DiD model is specified as follows:

$$y_{ijt} = \beta_0 + \beta_1\, Treatment_j \times AfterBan_t + \beta_2\, Treatment_j \times AfterLift_t + \eta_{jt} + \alpha_i + \gamma_t + \varepsilon_{ijt}$$

We use this model to estimate the effects of the ban and the subsequent lift on productivity, knowledge sharing, and skill acquisition among users in the treated country. For each user $i$ in country $j$ during week $t$, $y_{ijt}$ denotes the number of activities corresponding to each of the three dependent variables specified above. The variable $Treatment_j$ is a binary variable equal to 1 if country $j$ is the treatment country (Italy), and 0 otherwise. The variable $AfterBan_t$ equals 1 if week $t$ falls within the ban period–that is, after the ban and before the lift–and 0 otherwise. Similarly, $AfterLift_t$ equals 1 if week $t$ is after the lift, and 0 otherwise. Moreover, to further enhance the precision of our estimates, we include several control variables. We control for the number of working days $\eta_{jt}$ in each country $j$ during each week $t$, as national holidays may affect

---

the weekly activity levels of OSS developers. For the third model, where the dependent variable is skill acquisition, we also include several repository-level control variables to account for differences in project characteristics that may influence language adoption. In addition, we control for time-invariant user-level fixed effects $\alpha_i$ to account for individual characteristics that remain constant over time. Lastly, we include week-level fixed effects $\gamma_t$ to control for time-varying factors that may affect user activity across all countries, such as the end of a business quarter. Standard errors are clustered at user level in all three models. The average treatment effect of the ban on the treated group is captured by $\beta_1$, while the average treatment effect of the lift on the treated group is captured by $\beta_2$.

The DiD framework relies on the assumption of comparable composition between the treatment and control groups. While users in Italy, France, and Portugal exhibit similar macro-level activity patterns and the countries share geographic, economic, and regulatory proximity, differences in user composition may still exist. Such compositional differences can introduce bias into the estimation of treatment effects. To address this concern, we apply Propensity Score Matching (PSM) to construct a more comparable control group composed of users with similar characteristics to those in the treatment group during the pre-treatment period.[17] Specifically, we match users based on their profile information, activity data, and programming language usage observed prior to the ban.[18] Figure A3 in the Appendix provides a graphical representation of the reduction in bias across the matching. The balance between the matching variables before and after matching is presented in Table 3.

---

[17] We implement PSM using the radius matching method, imposing a common support condition and 0.001 caliper.
[18] To ensure feasibility and prevent overfitting, we limit matching to the top 50 programming languages, which cover over 93% of all languages used in the panel.

Table 3. Balance Table Before and After Matching

| Matching Variables | Mean | | | | t-test | | | |
|---|---|---|---|---|---|---|---|---|
| | Before Match | | After Match | | Before Match | | After Match | |
| | Treatment | Control | Treatment | Control | t | p > t | t | p > t |
| Has Company Affiliation Listed | 0.455 | 0.468 | 0.455 | 0.457 | -2.34 | 0.019 | -0.25 | 0.801 |
| Number of Public Repositories | 23.312 | 26.351 | 23.312 | 23.685 | -6.50 | 0.000 | -0.73 | 0.464 |
| Number of Public Gists | 2.486 | 2.801 | 2.486 | 2.449 | -1.70 | 0.089 | 0.17 | 0.863 |
| Number of Followers | 20.248 | 23.040 | 20.248 | 20.285 | -1.22 | 0.224 | -0.01 | 0.990 |
| Number of Followings | 17.093 | 20.332 | 17.093 | 16.425 | -0.46 | 0.646 | 0.12 | 0.907 |
| Tenure on Platform (Months) | 63.094 | 64.892 | 63.094 | 63.546 | -3.63 | 0.000 | -0.75 | 0.451 |
| Average Number of Code Commits | 3.437 | 3.234 | 3.437 | 3.167 | 2.00 | 0.045 | 2.07 | 0.039 |
| Average Number of Created Repositories | 0.208 | 0.200 | 0.208 | 0.202 | 2.19 | 0.029 | 1.16 | 0.245 |
| Average Number of Pull Requests | 0.210 | 0.286 | 0.210 | 0.220 | -6.47 | 0.000 | -0.91 | 0.365 |
| Average Number of Reviewed Pulls | 0.179 | 0.201 | 0.179 | 0.171 | -1.39 | 0.164 | 0.37 | 0.711 |
| Average Number of Opened Issues | 0.088 | 0.115 | 0.088 | 0.091 | -4.01 | 0.000 | -0.50 | 0.615 |
| Average Number of Initiated Discussions | 0.005 | 0.004 | 0.005 | 0.005 | 2.97 | 0.003 | 0.60 | 0.547 |
| Average Number of Joined Discussions | 0.002 | 0.002 | 0.002 | 0.002 | 0.45 | 0.652 | 0.26 | 0.794 |
| Average Number of Contributions to Private Repositories | 3.959 | 5.201 | 3.959 | 3.984 | -10.59 | 0.000 | -0.20 | 0.843 |

To verify whether the assumption of parallel trends holds across our three regression models, we employ a lead-lag specification following Autor (2003). In this approach, the eight weeks prior to the ban are noted as $t = -8, -7, \ldots, -1$, while the four weeks during the ban and the four weeks following the lift are denoted as $t = 1, 2, \ldots, 8$.[19] The model is specified as follows:

$$y_{ijt} = \beta_0 + \sum_{t=-n}^{n} \beta_1^{(t)} \, Treatment \times RelativeWeek_t + \eta_{jt} + \alpha_i + \gamma_t + \varepsilon_{ijt}$$

Following prior literature, we use the last period before the ban ($t = -1$) as the baseline. The variables follow our DiD model, with one key difference: instead of a single post-treatment indicator, we include a set of indicator variables for each week in the 16-week observation window. Specifically, for each week $t$, the variable $RelativeWeek_t$ equals 1 if the observation corresponds to week $t$, and 0 otherwise. The parallel trends assumption holds if the coefficients of the interaction terms ($\beta_1^{(t)}$) for the pre-treatment period are not statistically significant. We apply this model to the matched sample to test whether the parallel trends assumption is satisfied. The main analysis results, along with the parallel trends verification are presented in Section 5.

---

[19] We intentionally omit the notation $t = 0$ to enhance the clarity of the model.

## 5 Main Results

We now present the results of our analysis on the study's three main outcomes: productivity, knowledge sharing, and skill acquisition. Since we employed Poisson regression for this analysis (as described in Section 4.2), the coefficients are expressed as incidence rate ratios, with a baseline value of 1 indicating no effect.

### 5.1 Productivity

We begin by examining the impact of the ChatGPT ban on developer productivity within the DiD framework using the matched sample. As shown in Column (1) of Table 4, the DiD coefficient for the ban period–$Treatment \times AfterBan$–is less than one and statistically significant ($\beta_{DiD}^{(Ban)} = 0.936$, $p < 0.01$), indicating a 6.4% decrease in productivity (calculated as 0.936 – 1) during the ban period for developers in the treated country, Italy. This significant reduction in productivity in the absence of ChatGPT emphasizes the potential of LLMs to facilitate project initiation, enhance code development, and support overall project progression among OSS developers.

The $Treatment \times AfterLift$ coefficient captures the changes in productivity after the ban was lifted, relative to the pre-ban period in Italy. We do not observe any significant impact on productivity among users in Italy following the lifting of the ban, as indicated by the statistically insignificant coefficient on the corresponding term. This return to pre-ban levels is anticipated, as reinstating access to ChatGPT does not necessarily lead to further gains beyond those already realized prior to the ban.

## 5.2 Knowledge Sharing

Next, we examine the impacts of the ChatGPT ban and its subsequent lifting on knowledge sharing among users in the treated group. As shown in Column (2) of Table 4, the DiD coefficient for the ban period–$Treatment \times AfterBan$–is not statistically significant at conventional significance levels. This means that we cannot reject the null hypothesis that the ban on ChatGPT had no effect on the knowledge sharing of users in Italy compared to pre-ban activity levels.

However, unlike the impact of lifting the ban on productivity, the lift seems to lead to an increase in the knowledge sharing activity levels among users in the treated group. As shown in Table 4, the DiD coefficient for the after lift period–$Treatment \times AfterLift$–is greater than one and statistically significant ($\beta_{DiD}^{(Lift)} = 1.096$, $p < 0.05$). This indicates a 9.6% increase in knowledge-sharing activity (calculated as 1.096 – 1) after the lift period relative to the pre-ban period among users in the treated country, Italy. It appears that with reinstated access to ChatGPT, developers in Italy may have experienced a "released capacity" stemming from the benefits of this tool and directed a portion of this capacity toward engaging more in collaborative knowledge-sharing activities.

## 5.3 Skill Acquisition

Lastly, we examine the impacts of the ChatGPT ban and its subsequent lifting on skill acquisition among the treated group of users. As shown in Column (3) of Table 4, the DiD coefficient for the ban period–$Treatment \times AfterBan$–is less than one and significant ($\beta_{DiD}^{(Ban)} = 0.916$, $p < 0.1$), indicating an 8.4% decrease in skill acquisition (calculated as 0.916 – 1) relative to the pre-ban period among users in the treated country, Italy. This finding suggests that access to LLMs like ChatGPT plays a meaningful role in facilitating developers' learning, particularly their ability to adopt new programming languages.

The lifting of the ban does not have a statistically significant impact on skill acquisition of the treatment group. The DiD coefficient for the after-lift period–$Treatment \times AfterLift$–is not statistically significant at conventional significance levels, indicating no measurable change in skill acquisition following the restoration of access to ChatGPT.

**Table 4. Main Results**

| Variables | (1) Productivity | (2) Knowledge Sharing | (3) Skill Acquisition |
|---|---|---|---|
| $Treatment \times AfterBan$ | 0.936*** | 0.972 | 0.916* |
|  | (0.021) | (0.035) | (0.047) |
| $Treatment \times AfterLift$ | 0.993 | 1.096** | 1.013 |
|  | (0.026) | (0.050) | (0.054) |
| Working Days | 1.069*** | 1.073*** | 0.999 |
|  | (0.018) | (0.027) | (0.045) |
| Constant | 10.375*** | 2.787*** | 0.092*** |
|  | (0.853) | (0.336) | (0.028) |
| Observations | 298,632 | 132,598 | 70,766 |
| User FE | YES | YES | YES |
| Week FE | YES | YES | YES |
| Pseudo $R^2$ | 0.507 | 0.535 | 0.479 |

**Note:** The dependent variables are Productivity, Knowledge Sharing, and Skill Acquisition in columns (1), (2), and (3). The DiD coefficients represent the incidence rate ratio, with 1 being the baseline. The results for the additional control terms are omitted to maintain brevity. Robust standard errors are clustered by User ID and are reported in parentheses. The significance levels are denoted by ***p < 0.01, **p < 0.05, *p < 0.1.

## 5.4 Pre-Ban Trends

To assess the validity of the parallel trends assumption, we employ the lead-lag model outlined in Section 4.2. The estimation results are presented in Table 5, with a visual representation of the corresponding coefficients shown in Figures 4 through 6 in the Appendix. None of the pre-treatment coefficients are statistically significant, providing support for the parallel trends assumption

**Table 5. Pre-Treatment Trends**

| Relative Week | (1)<br>**Productivity** | (2)<br>**Knowledge Sharing** | (3)<br>**Skill Acquisition** |
|---|---|---|---|
| Relative Week (-8) | 1.060<br>(0.047) | 1.100<br>(0.077) | 0.961<br>(0.096) |
| Relative Week (-7) | 1.066<br>(0.046) | 0.998<br>(0.070) | 0.877<br>(0.088) |
| Relative Week (-6) | 1.013<br>(0.043) | 1.025<br>(0.068) | 1.115<br>(0.107) |
| Relative Week (-5) | 1.030<br>(0.041) | 0.916<br>(0.065) | 1.000<br>(0.106) |
| Relative Week (-4) | 0.994<br>(0.038) | 0.991<br>(0.067) | 1.051<br>(0.099) |
| Relative Week (-3) | 0.986<br>(0.037) | 1.046<br>(0.065) | 0.908<br>(0.085) |
| Relative Week (-2) | 0.992<br>(0.036) | 0.946<br>(0.053) | 1.164<br>(0.116) |
| Relative Week (-1) | Omitted | Omitted | Omitted |
| Relative Week (+1) | 0.921**<br>(0.032) | 0.898<br>(0.062) | 0.938<br>(0.093) |
| Relative Week (+2) | 0.971<br>(0.038) | 1.036<br>(0.065) | 0.901<br>(0.096) |
| Relative Week (+3) | 0.993<br>(0.039) | 0.982<br>(0.057) | 0.965<br>(0.109) |
| Relative Week (+4) | 0.926*<br>(0.041) | 0.990<br>(0.071) | 0.876<br>(0.103) |
| Relative Week (+5) | 0.987<br>(0.039) | 1.033<br>(0.065) | 0.998<br>(0.104) |
| Relative Week (+6) | 0.980<br>(0.041) | 1.058<br>(0.067) | 0.946<br>(0.109) |
| Relative Week (+7) | 1.006<br>(0.045) | 1.209***<br>(0.081) | 1.098<br>(0.130) |
| Relative Week (+8) | 1.058<br>(0.049) | 1.106<br>(0.070) | 1.077<br>(0.111) |
| Observations | 298,632 | 132,598 | 70,766 |
| User FE | YES | YES | YES |
| Week FE | YES | YES | YES |
| Pseudo $R^2$ | 0.507 | 0.536 | 0.479 |

**Note:** Periods before the ban are indicated with a negative sign, while periods after the ban are indicated with a positive sign. The notation does not include a period denoted as 0 for ease of interpretation. The DiD coefficients represent the incidence rate ratio with 1 being the baseline. Robust standard errors are clustered by User ID and are reported in parentheses. The results for the additional control terms in the third column are omitted to maintain brevity. The significance levels are denoted by ***p < 0.01, **p < 0.05, *p < 0.1.

# 6 Robustness Checks

To ensure the reliability and consistency of our analysis, we conduct additional robustness checks and examine and discuss potential threats to identification. First, we examine whether developers in the treated country may have switched to alternative general-purpose LLMs (e.g., Google Gemini, Llama, Claude) following the ChatGPT ban. Second, we assess the possible substitution with the programming-specific LLM available at the time, GitHub Copilot. Lastly, we investigate whether developers in the treated country may have used circumvention tools such as VPNs to bypass the ban.

First, to investigate the possibility of substitution to alternative general-purpose LLMs, we consider the timeframe of our study: February 4, 2023, to May 26, 2023. During this period, no alternative general-purpose LLMs were publicly available in either the treatment or control countries. We verified this by reviewing the release timelines of the three major general-purpose LLMs: Google Bard (now Gemini) became available in Europe on July 13, 2023;[20] Meta's Llama was publicly released on July 18, 2023;[21] and Anthropic's Claude was released in Europe in May 2024.[22] Since all of these launches occurred after the end of our observation window, developers in the treated country could not have possibly switched to alternative general-purpose LLMs during the period under study. Therefore, our analysis and results are robust to concerns about substitution to other general-purpose LLMs during the ban.

Next, we investigate the possibility that developers in the treated country substituted GitHub Copilot (GHC) for ChatGPT during the ban period.[23] To assess this potential substitution,

---

[20] https://techcrunch.com/2023/07/13/googles-bard-finally-lands-in-the-eu-now-supports-more-than-40-languages
[21] https://ai.meta.com/llama/license/
[22] https://www.france24.com/en/europe/20240514-anthropic-launches-ai-assistant-claude-in-europe-following-us-debut
[23] It is important to note that GitHub Copilot is a paid service. Given the voluntary nature of participation in open-source software (OSS) development and the prior availability of ChatGPT as a free tool, it is unlikely that a significant

we split the data into two mutually exclusive subsamples: (1) GHC-covered repositories–projects written exclusively in languages fully supported by GHC,[24] and (2) Non-GHC-covered repositories– projects written entirely in languages that GHC does not support. If developers widely substituted ChatGPT with GHC, the treated group should have been able to offset some of the ban's effects when working in GHC-covered repositories. We would therefore expect smaller treatment effects—on productivity, knowledge sharing, and skill acquisition—in the GHC-covered subsample than in the non-GHC-covered subsample. Conversely, if the estimated effects are similar across both subsamples, it would indicate that meaningful substitution to GHC did not occur.

Using the empirical framework described in Section 4.2, we re-estimated the models separately for the two subsamples—GHC-covered and non-GHC-covered repositories. The results, reported in Table 6,[25] show similar treatment effects on productivity, knowledge sharing, and skill acquisition across both subsamples and are consistent with the baseline estimates. Because the magnitude and direction of the effects do not differ meaningfully between GHC-covered and non-GHC-covered projects, we find no evidence that Italian developers substantially substituted GitHub Copilot for ChatGPT during the ban. These findings strengthen the robustness of our main results and indicate that potential switching to GitHub Copilot does not compromise the study's conclusions.

---

number of OSS contributors would opt to pay for an alternative tool in response to the restriction. Regardless, we'll test this assumption in this section.

[24] Even if a language wasn't fully supported by GHC, its large training dataset may have still enabled it to generate some (possibly sub-optimal) code in that language.

[25] The two subsamples do not sum to the full dataset because (i) projects that mix GHC-supported and unsupported languages are dropped from both subsamples, and (ii) the language filter creates singleton fixed-effect cells that are automatically omitted in estimation. Hence, the combined observation count is smaller than in the original data.

**Table 6. Regression Results for the GHC vs. Non-GHC Covered Repositories**

| | Productivity | | | Knowledge Sharing | | | Skill Acquisition | | |
|---|---|---|---|---|---|---|---|---|---|
| Sample Construction | (1) GHC Repos | (2) Non-GHC Repos | (3) Main Results | (4) GHC Repos | (5) Non-GHC Repos | (6) Main Results | (7) GHC Repos | (8) Non-GHC Repos | (9) Main Results |
| $Treatment \times AfterBan$ | 0.948** (0.024) | 0.915** (0.041) | 0.936*** (0.021) | 0.972 (0.035) | 0.982 (0.153) | 0.972 (0.035) | 0.838** (0.062) | 0.913 (0.091) | 0.916* (0.047) |
| $Treatment \times AfterLift$ | 0.999 (0.031) | 0.988 (0.050) | 0.993 (0.026) | 1.058 (0.046) | 1.235 (0.251) | 1.096** (0.050) | 0.942 (0.079) | 1.069 (0.109) | 1.013 (0.054) |
| Working Days | 1.029* (0.018) | 1.082** (0.037) | 1.069*** (0.018) | 1.060** (0.028) | 1.091 (0.087) | 1.073*** (0.027) | 1.038 (0.074) | 0.910 (0.074) | 0.999 (0.045) |
| Constant | 14.848*** (1.221) | 8.764*** (1.453) | 10.375*** (0.853) | 3.584*** (0.457) | 1.628 (0.600) | 2.787*** (0.336) | 0.057*** (0.026) | 0.166*** (0.081) | 0.092*** (0.028) |
| Observations | 149,410 | 108,043 | 298,632 | 79,373 | 18,690 | 132,598 | 29,428 | 26,833 | 70,766 |
| User FE | YES | YES | YES | YES | YES | YES | YES | YES | YES |
| Week FE | YES | YES | YES | YES | YES | YES | YES | YES | YES |
| Pseudo $R^2$ | 0.477 | 0.567 | 0.507 | 0.517 | 0.467 | 0.535 | 0.508 | 0.409 | 0.479 |

**Note:** The dependent variable is Productivity in columns (1), (2), and (3), Knowledge Sharing in columns (4), (5), and (6), and Skill Acquisition in columns (7), (8), and (9). The DiD coefficients represent the incidence rate ratio with 1 being the baseline. The results for the additional control terms in the third column are omitted to maintain brevity. Robust standard errors are clustered by User ID and are reported in parentheses. The significance levels are denoted by ***p < 0.01, **p < 0.05, *p < 0.1.

Lastly, we consider the potential use of circumvention tools, such as Virtual Private Networks (VPNs), by developers in the treated country during the ChatGPT ban. VPN use could allow some users to bypass the ban, thereby introducing untreated users into the treated group. This would attenuate the estimated treatment effects in our DiD framework, leading only to an underestimation of ChatGPT's true impact. To address this concern, we conduct a supplementary analysis by restricting our observation window to the first week of the ban, when widespread VPN adoption was likely minimal.

As detailed in Appendix Section A2, the results from this restricted sample are both qualitatively and quantitatively consistent with our main findings. The estimated effects on productivity, knowledge sharing, and skill acquisition remain stable when limiting the observation window to only the first week after the ban. This suggests that VPN usage did not materially bias our estimates, reinforcing the validity of our conclusions.

## 7 Heterogenous Treatment Effects on Developers

In the preceding sections, we analyzed the average treatment effect (ATT) of the ChatGPT ban and its subsequent lifting on productivity, knowledge sharing, and skill acquisition among all users in the treatment group. In this section, we explore whether these effects vary across different developer subgroups. By examining how the unavailability of ChatGPT affected users with differing characteristics—differentiated by factors such as experience—we aim to uncover heterogenous patterns that may not be evident in the aggregate data.

One important factor influencing productivity, knowledge sharing, and skill acquisition in the OSS community is user experience, which we proxy using the tenure of the user on the platform.[26] Following the literature (Cui et al. 2024), we hypothesize that the benefits users derive from using LLMs vary with their experience level. The rationale behind this idea is that more experienced users, becoming more proficient over time, may rely less on and benefit less from ChatGPT. To test this hypothesis, we divide our sample into three subgroups based on tenure: (1) *novice* users, with tenure in the bottom 25th percentile; (2) *intermediate* users, with tenure between the 25th and 75th percentiles; (3) *advanced* users, with tenure in the top 25th percentile.

We estimate the impact of the ban and lift productivity, knowledge sharing, and skill acquisition for each experience group by applying the empirical framework outlined in Section 4.2 to each corresponding sub-sample.

## 7.1 Productivity

The results of the heterogeneity analysis for the productivity variable are shown in Table 7. As shown in columns (1) to (3), the ban on ChatGPT led to a 15.2% decrease (calculated as 0.848 – 1) in productivity for novice users in the treated country, which they did not fully recover from in the four weeks following the lift. Novice users appear to be the most adversely affected by the loss of access to ChatGPT. This may be due to novice users' over-reliance on the tool or a lack of deep expertise to sustain pre-ban productivity levels independently. Their incomplete recovery four weeks post-lift suggests a lasting disruption to their workflows.

In contrast, intermediate and advanced users did not experience significant changes in productivity due to the ban. Reinstating access to ChatGPT after lifting the ban did not significantly affect productivity for any subgroup relative to the pre-ban period. This may be because more experienced users relied less on ChatGPT or possessed sufficient expertise to adapt without it.

---

[26] We define user tenure on GitHub as the number of months since the user joined the platform.

These findings highlight the crucial role that LLMs like ChatGPT can play in supporting less experienced developers.

**7.2 Knowledge Sharing**

The results of the heterogeneity analysis on the knowledge sharing variable are presented in Table 7, columns (4) to (6). The ban on ChatGPT did not significantly impact knowledge sharing across any experience group. However, the resumption of access led to a 22.3% increase (calculated as 1.223 – 1) in knowledge-sharing activity among intermediate users, with no significant changes observed for novice or advanced users. This indicates that intermediate users are particularly well-positioned to leverage ChatGPT for reviewing peer contributions, participating in discussions, and sharing insights. They may have enough experience to use advanced tools effectively while still actively seeking collaborative learning opportunities, making them especially responsive to ChatGPT's return.

On the other hand, the ban and its subsequent lifting did not affect either novice or advanced users, which may be due to different underlying factors. Novice users may not yet have the proficiency to fully exploit ChatGPT for knowledge sharing, possibly due to limited experience or confidence in contributing to discussions–as suggested by the limited number of such activities in Column (4) of Table 7. Advanced users, however, might rely more on their own expertise and established networks, making them less dependent on AI tools for knowledge sharing. These findings suggest that LLMs like ChatGPT can play a significant role in facilitating collaborative learning and information exchange, particularly among software developers at intermediate experience levels.

**7.3 Skill Acquisition**

Lastly, the results of the heterogeneity analysis on the skill acquisition variable are presented in Table 7, columns (7) to (9). The ban on ChatGPT led to a 15.2% decrease (calculated as 0.848 – 1) in skill acquisition among intermediate users in the treated country. This indicates that intermediate users may rely on ChatGPT as a valuable resource for learning new skills and improving their development practices, consistent with the positive effect observed on their knowledge sharing. Intermediate users' reliance may be attributed to their position along the learning curve, where they possess enough foundational knowledge to effectively use LLMs, yet continue to seek opportunities to broaden their skill sets.

In contrast, novice and advanced users did not experience significant changes in skill acquisition due to either the ban or the lift, relative to the pre-ban period. Novice users may not yet fully leverage ChatGPT for skill acquisition, as they are likely still focused on mastering foundational concepts through traditional learning methods. Advanced users, on the other hand, may rely more on their extensive experience and less on LLMs for acquiring new skills. These findings indicate that LLMs like ChatGPT can play a particularly important role in supporting skill development among intermediate users.

The analysis provided in this section, offers important insights into how different users in the OSS community benefit from using LLMs such as ChatGPT. Novice users experience a significant decline in productivity when they lose access to ChatGPT, as evidenced by a 14.8% reduction in productivity. Intermediate users also experience losses, albeit in a different dimension: their skill acquisition drops by 15.2% after the ban. Moreover, once access to ChatGPT is reinstated, intermediate users show a 22.3% increase in knowledge sharing, possibly because access to ChatGPT helps reduce the cognitive load associated with engaging in collaborative activities, making it easier to review code, provide feedback, and engage in discussions. These findings suggest that different user groups benefit from ChatGPT in distinct ways. While novice users primarily gain in terms of productivity, intermediate users leverage it in ways that extend beyond mere productivity gains; to share knowledge across the OSS community and learn new skills along the way.

## Table 7. Regression Results by Developer Experience Level

|  | Productivity | | | Knowledge Sharing | | | Skill Acquisition | | |
|---|---|---|---|---|---|---|---|---|---|
|  | (1) | (2) | (3) | (4) | (5) | (6) | (7) | (8) | (9) |
| Variable | Novice | Intermediate | Advanced | Novice | Intermediate | Advanced | Novice | Intermediate | Advanced |
| $Treatment \times AfterBan$ | 0.848*** | 0.950 | 1.004 | 0.852 | 1.033 | 0.954 | 1.048 | 0.848** | 0.891 |
|  | (0.036) | (0.030) | (0.043) | (0.120) | (0.064) | (0.043) | (0.082) | (0.063) | (0.115) |
| $Treatment \times AfterLift$ | 0.915* | 0.991 | 1.071 | 0.950 | 1.223** | 1.030 | 1.138 | 0.945 | 0.816 |
|  | (0.045) | (0.038) | (0.057) | (0.145) | (0.099) | (0.058) | (0.100) | (0.075) | (0.117) |
| Working Days | 1.047 | 1.065** | 1.100*** | 1.091 | 1.003 | 1.125*** | 1.009 | 0.993 | 0.989 |
|  | (0.040) | (0.027) | (0.033) | (0.078) | (0.041) | (0.039) | (0.072) | (0.069) | (0.116) |
| Constant | 12.593*** | 9.575*** | 9.871*** | 1.629 | 2.950*** | 2.941*** | 0.123*** | 0.107*** | 0.048*** |
|  | (2.263) | (1.149) | (1.407) | (0.563) | (0.570) | (0.482) | (0.067) | (0.046) | (0.032) |
| Observations | 64,612 | 148,961 | 85,059 | 14,396 | 62,674 | 55,528 | 28,123 | 30,268 | 12,375 |
| User FE | YES | YES | YES | YES | YES | YES | YES | YES | YES |
| Week FE | YES | YES | YES | YES | YES | YES | YES | YES | YES |
| Pseudo $R^2$ | 0.46 | 0.49 | 0.57 | 0.40 | 0.49 | 0.59 | 0.45 | 0.49 | 0.56 |

**Note:** The dependent variable is Productivity in columns (1), (2), and (3), Knowledge Sharing in columns (4), (5), and (6), and Skill Acquisition in columns (7), (8), and (9). The p-values for the test of difference among the coefficients of ban and lift are 0.032 and 0.394 for productivity, 0.339 and 0.077 for knowledge sharing, 0.055 and 0.304 for skill acquisition. The DiD coefficients represent the incidence rate ratio with 1 being the baseline. The results for the additional control terms in the third model are omitted to maintain brevity. Robust standard errors are clustered by User ID and are reported in parentheses. The significance levels are denoted by ***p < 0.01, **p < 0.05, *p < 0.1.

# 8 Impact on Learning Different Programming Languages

The value of large language models (LLMs) in skill acquisition is not uniform—it varies with the nature of the task. In domains where learning is hindered by fragmented resources or rapid changes, LLMs can serve as crucial complements by offering real-time, tailored support. This section examines how LLMs facilitate learning across different programming language categories, revealing their role in helping developers overcome task-specific barriers.

We examine this variation through programming languages, where learning demands differ by context and complexity. Programming languages serve diverse goals and use cases, including areas such as web development, general-purpose programming, and scientific computing. The importance of LLMs in supporting programming language learning varies significantly across different contexts. This variation depends on factors such as the availability of alternative learning resources or *outside options* (e.g., structured documentation, online tutorials, and community support) as well as the inherent complexity of the language itself. For languages with abundant, well-organized resources, learners may experience minimal differences in learning outcomes with or without LLM access. In contrast, for languages with limited to poorly structured resources, the absence of LLM support can create significant learning barriers, making LLMs particularly valuable in these contexts.

To explore this heterogeneous impact, we focus on the top 50 programming languages used across user projects in the panel. Collectively, these languages account for over 93% of the programming languages employed in these projects.[27] We then categorize them into seven major clusters based on their primary applications.[28,29] These clusters are summarized in Table 8 and further detailed in Section A3 of the Appendix.

---

[27] A detailed list of these languages and their relative usage frequencies is presented in Figure A7 of the Appendix.

**Table 8. Programming Language Clusters by Application**

| Application | Purpose | Programming Languages |
|---|---|---|
| General-Purpose | Versatile languages used for a wide range of programming tasks, from application development to automation. | Java, Python, C#, Go, Perl, Kotlin |
| Web Development | Languages and technologies specialized for creating and managing web applications, covering both front-end and back-end tasks, including browser rendering and HTTP communication. | JavaScript, HTML, CSS, TypeScript, SCSS, PHP, Ruby, Vue.js, Blade, Dart, Twig, Smarty, Handlebars.js, Less.js, Svelte, Dart |
| System Programming | Languages used to develop low-level software such as operating systems, device drivers, and other system utilities that require direct interaction with hardware. | C, C++, Rust, Assembly |
| Scientific Computing | Languages and technologies for numerical computation, data analysis, and mathematical modeling, commonly used in fields like engineering, physics, and data science. | MATLAB, R |
| DevOps and Configuration | Languages and technologies used for infrastructure management, automation, and orchestration, enabling infrastructure as code (IaC) and deployment pipelines. | Dockerfile, Makefile, CMake, Batchfile, Nix, Procfile, PowerShell, Starlark |
| Templating and Markup | Languages and technologies used for structuring and styling content, typically used to generate dynamic content while separating content from design. | TeX, Roff, XSLT, Jinja |
| Domain-Specific | Specialized languages and technologies used for particular industries, devices, or narrowly defined tasks requiring customized functionality. | Objective-C++, Solidity, Hack, Swift, Objective-C, Lua, HCL, GLSL, m4 |

**Note:** While many programming languages are versatile and can be applied across various domains, the clustering presented here focuses on their most widely adopted and primary use cases. We acknowledge that many of these technologies and tools can be used in other domains beyond their designated categories; but for clarity and analytical consistency, each has been grouped according to its predominant application.

---

[28] It is important to note that while many programming languages are versatile and applicable across multiple domains, each category in this classification reflects the most common and specialized use of a given language. This categorization is based on the primary, most relevant applications for which these languages are typically used.

[29] We exclude Jupyter Notebook from our analysis, as it is an interactive environment rather than a standalone programming language, and supports multiple languages (e.g., Python, R). Its underlying language cannot be reliably identified from repository metadata, making accurate classification infeasible.

To assess the impact of the ban and its lift on the learning rate of programming languages in each of the seven categories, we construct dependent variables representing the number of new programming languages from each cluster that a user employed in their projects. For estimation, we use the same empirical framework outlined in Section 4.2. We further include additional repository-level control variables to improve estimate precision. These repository-level control variables include the number of stars, forks (copies), opened issues, and indicators for whether the repository is a fork from a master repository, whether it has open issues, a projects page, downloads, a Wikipedia page, an information page, and discussions. We also control for repository size (in lines of code), the total number of programming languages used, and include interaction terms and their exponentials to account for potential non-linear relationships. Lastly, based on our findings in Section 7.3—that only intermediate users benefit from LLMs for skill acquisition—we restrict this analysis to this subgroup.[30] The results of this regression are presented in Table 9.

**Table 9. New Programming Language Acquisition by Cluster for Intermediate Users**

|  | (1) General Purpose | (2) Web Development | (3) System Programming | (4) Scientific Computing | (5) DevOps and Configuration | (6) Templating and Markup | (7) Domain-Specific |
|---|---|---|---|---|---|---|---|
| $Treatment \times AfterBan$ | 1.202 (0.219) | 0.692* (0.103) | 0.499** (0.121) | 2.215 (1.545) | 0.867 (0.141) | 1.664 (0.650) | 0.355** (0.113) |
| $Treatment \times AfterLift$ | 1.193 (0.232) | 0.743 (0.131) | 0.593 (0.172) | 1.229 (0.936) | 0.949 (0.185) | 1.763 (0.723) | 0.794 (0.226) |
| Obs. | 10,132 | 15,450 | 6,027 | 1,401 | 11,100 | 3,773 | 6,322 |
| User FE | YES | YES | YES | YES | YES | YES | YES |
| Week FE | YES | YES | YES | YES | YES | YES | YES |
| Mean | 0.262 (0.504) | 0.419 (0.853) | 0.285 (0.588) | 0.133 (0.354) | 0.291 (0.584) | 0.183 (0.431) | 0.224 (0.530) |
| Pseudo $R^2$ | 0.22 | 0.38 | 0.31 | 0.40 | 0.33 | 0.39 | 0.40 |

Note: Column (1) reports the dependent variable as the number of new programming languages learned within the General-Purpose category. Columns (2) through (7) report the same for Web Development, System Programming, Scientific Computing, DevOps and Configuration, Templating and Markup, and Domain-Specific categories, respectively. Control variables are omitted for brevity. The DiD coefficients represent the incidence rate ratios (IRRs), with 1 being the baseline. Robust standard errors are clustered at the User ID level and are reported in parentheses. The significance levels are denoted by ***p < 0.01, **p < 0.05, *p < 0.1.

---

[30] We also conducted the same analysis on the full sample, as described in Section A4 of the Appendix. We find no significant effect of the ban or its lift on learning any language cluster, except for a 29.8% decrease in domain-specific language learning.

As shown in columns (2), (3), and (7) of Table 9, the loss of access to ChatGPT led to a 30.8% decrease (calculated as 0.692 – 1) in learning web development languages, a 50.1% decrease (calculated as 0.499 – 1) in learning system programming languages, and a 64.5% decrease (calculated as 0.355 – 1) in learning domain-specific languages.[31] These findings suggest that users may have been particularly reliant on ChatGPT to overcome the complexities associated with learning programming languages and technologies within these three clusters.

Based on these results, ChatGPT likely played a crucial role in helping developers overcome key barriers to learning within these three clusters. In the web development cluster, skill acquisition often requires integrating multiple languages (e.g., HTML, CSS, JavaScript, and TypeScript) across both front-end and back-end environments (Mokoginta et al. 2024; Piastou 2023). This process requires not only understanding each language individually but also their interactions within dynamic and complex application structures. Moreover, many challenges in web development are visual or layout-related (Arab et al. 2025), where identifying and correcting issues without real-time feedback can be difficult. ChatGPT likely supported users in navigating these complexities by providing cross-language explanations, rapid prototyping assistance, and real-time debugging support, thereby making the learning process significantly more accessible.

In systems programming, developers must understand foundational system-level concepts like memory management, pointer arithmetic, and concurrent processing. These tasks often involve navigating cryptic compiler and runtime errors, including segmentation faults and borrow-checker violations (Lee et al. 2018; Oorschot et al. 2023).[32] Traditional learning resources for these topics are often highly technical and assume substantial prior knowledge, making self-directed learning particularly difficult. ChatGPT likely functioned as an on-demand tutor, translating vague error messages into actionable guidance and breaking down abstract concepts into manageable steps.

---

[31] While these estimated percentage decreases are large in magnitude, it is important to note that the mean values of the dependent variables in these analyses are relatively low. Therefore, these results should be interpreted primarily in terms of the direction of impact and the relative differences across language clusters, rather than their absolute magnitudes.

[32] A borrow-checker violation happens when code breaks memory safe access rules, such as using data after it's been moved or accessed by multiple owners at the same time.

Similarly, domain-specific languages like Solidity or Objective-C++ present unique learning challenges due to fragmented documentation (Kannengießer et al. 2022), rapid and frequent version changes (Mitropoulos et al. 2024), and domain-specific complexity (Peng et al. 2025). In such cases, ChatGPT may have filled knowledge gaps by providing accessible guidance in specialized environments where conventional community support is often sparse. Together, these findings highlight the particular value of LLMs like ChatGPT in supporting skill acquisition in technically complex, fragmented, or rapidly evolving areas of software development.

## 9 Conclusion and Discussion

### 9.1 Conclusion

Our study examines the multifaceted impacts of LLMs on a subgroup of knowledge workers–software developers. Using a large-scale natural experiment stemming from the temporary ban of ChatGPT in Italy, we collect and analyze developer activity data from Italy, France and Portugal. Specifically, we use the negative outcomes associated with losing access to ChatGPT–the world's most widely used LLM–as a proxy to infer that access to LLMs enhances productivity, knowledge sharing, and skill acquisition within GitHub, the world's largest OSS community.

In addition, we show that LLMs can provide heterogeneous support in learning and employing different types of programming languages. Our results suggest that access to ChatGPT significantly improves the learning rate of more complex and rapidly evolving programming languages, such as those used in web development, or areas requiring highly specialized knowledge such as domain-specific programming languages. By contrast, in more general fields where there is relatively easy access to alternative learning resources such as official documentation and active online communities, developers do not benefit significantly from the learning support provided by LLMs and can instead rely on traditional learning methods.

Our study advances the literature on the business value of AI by examining the multi-dimensional, sustained effects of LLMs on an important subgroup of knowledge workers–software developers–beyond mere productivity impacts. We broaden the scope of prior research, which has largely focused on isolated, temporary question-answer interactions, by showing how LLMs shape ongoing, structured collaborative work. Specifically, we demonstrate that LLMs enhance both collaborative knowledge exchange and skill acquisition across OSS developers, providing benefits that extend beyond routine task automation. Furthermore, we extend the literature on LLMs' impact on various worker subgroups by uncovering heterogeneous effects based on worker experience. We show that while access to LLMs significantly enhances knowledge sharing and skill acquisition for intermediate users, novice users appear heavily dependent on it for maintaining productivity—so much so that losing access results in persistent performance declines even after access is restored. Lastly, we extend the literature on context-dependent complementarity in LLM-assisted learning. Rather than treating skill acquisition as uniform, we show that the benefits of LLMs vary by context type—providing the greatest support for learning programming languages that are technically complex, poorly documented, or rapidly evolving. We argue that LLMs serve as valuable complements in cognitively demanding learning contexts where traditional resources fall short. Collectively, these contributions provide novel empirical evidence on how LLMs reshape structured collaborative work by examining three interrelated dimensions of productivity, knowledge sharing, and skill acquisition across varying user skill levels.

**9.2 Limitations**

Our study must be viewed within the boundaries of its limitations. First, we do not observe which LLMs the users in our panel are using. Nevertheless, given ChatGPT's dominance and superior performance during the early phase of the LLM boom (post-ChatGPT 3.5 launch), as well as the unavailability of alternative general-purpose LLMs during the study timeframe (as shown in Section 6), this limitation is unlikely to introduce significant bias.

Second, some users may have employed circumvention tools, such as VPNs, to access ChatGPT during the ban period. We address this concern in Section 6 and empirically demonstrate that such behavior has no significant impact on our estimates. We argue that, to the extent VPN use did occur, it would more likely dilute the observed effect, leading to an underestimation of the impact's magnitude rather than an overestimation.

Lastly, there is a possibility that the developers in the treated country may have substituted ChatGPT with GitHub Copilot (GHC), a programming-specific LLM, during the ban period. However, our empirical analysis in Section 6 directly addresses this possibility by comparing outcomes between GHC-covered and non-GHC-covered repositories—i.e., those written entirely in languages fully supported by GHC versus those exclusively written in unsupported languages. If developers had broadly adopted GHC as a replacement, we would observe weaker treatment effects in GHC-covered repositories. Yet, our findings reveal consistent patterns across both groups, indicating that substantial substitution to GHC is unlikely. Furthermore, if any such substitution did occur, it would likely reduce the observed contrast between treated and control groups. This implies that in this unlikely scenario, our estimates may in fact understate the true effect of the ChatGPT ban, rather than overstate it.

### 9.3 Managerial and Policy Implications

Our study has several major managerial and policy implications. We argue that developer productivity is a multidimensional construct shaped not only by direct code contributions but also by broader processes such as onboarding, learning, and knowledge sharing and show that the impact of LLMs on productivity extends far beyond immediate code development assistance. While conventional productivity metrics capture short-term returns, they overlook the broader, compounding effects that LLMs generate through improved onboarding, accelerated learning, and richer knowledge exchange. By lowering barriers to entry, supporting real-time skill acquisition, and facilitating peer knowledge exchange, LLMs reshape the very foundations of both developer and organizational productivity. These indirect, ecosystem-wide benefits accumulate over time, which strengthen the collective output and agility of the organization. In highlighting these multidimensional effects, our research offers a more holistic and forward-looking understanding of how LLMs transform productivity in software development.

By showing the differentiated impacts of LLMs across user experience levels, we demonstrate the importance of customized workforce training and support strategies. Novice developers particularly benefit from productivity enhancements through AI-driven tools, indicating that targeted AI adoption programs can accelerate the onboarding process and quickly increase junior developers' productivity and contributions. Moreover, by lowering the barriers to entry, LLMs enable new developers to engage with complex codebases and workflows earlier in their tenure, allowing them to make more meaningful contributions from the outset. Leveraging LLMs also presents an opportunity for firms to diversify and expand their talent pipeline by more effectively integrating junior developers with varied technical backgrounds. Finally, by minimizing the need for constant supervision or intensive mentoring, AI-driven support systems help organizations scale their development teams more efficiently during periods of growth.

Moreover, drawing from the differentiated impacts of LLMs across user experience levels, we show that intermediate-level developers uniquely leverage LLMs to enhance knowledge sharing across the community. Managers should thus encourage the use of LLMs among mid-level developers to facilitate more effective collaborative environments, improved problem-solving, and richer knowledge exchange within teams. Beyond these immediate benefits, empowering intermediate developers to share knowledge through LLMs can strengthen long-term knowledge retention, reduce reliance on single experts, and improve organizational resilience. Together, these dynamics can further enhance both intra- and inter-team knowledge exchange, ultimately contributing to enhanced organizational productivity.

In addition, our findings suggest a substantial organizational benefit of LLMs from the perspective of workforce training and skill development. The ability of mid-level employees to rapidly upskill using LLMs enables organizations to respond more effectively to technological changes and evolving business needs. By supporting self-directed, on-demand learning, LLMs shift the paradigm of workforce development away from static training programs toward more dynamic, contextualized, and scalable models of skill acquisition, aligned with the workers' experience level. This shift broadens the internal talent pool, facilitating lateral mobility and enabling companies to deploy human capital more flexibly in response to shifting priorities. Furthermore, this enhanced learning capacity fosters greater organizational strategic agility, allowing companies to adapt, innovate, and maintain competitive advantage in fast-changing environments. Together, these dynamics not only cultivate a more resilient and future-ready workforce but also generate positive spillover effects on productivity by amplifying both individual contributions and collective performance across the organization.

Beyond capability-building and workforce enablement, our findings offer critical insight into the organizational risks associated with LLM service interruptions.[33] We show that while intermediate users tend to recover quickly once access is restored, novice developers experience persistent productivity declines even after the disruption ends. This asymmetry underscores the importance of organizational resilience planning in AI adoption. Specifically, firms must account not only for the benefits of LLM integration but also for the potential costs of service outages—particularly for less experienced employees who rely more heavily on AI support. As organizations increasingly depend on third-party LLM providers,[34] these findings call for a more deliberate LLM strategy—one that aligns infrastructure decisions with workforce composition. Managers may consider investing in more stable vendors, diversifying provider dependencies, or developing internal LLM capabilities to mitigate service risk.

Finally, our findings carry important policy implications for governments and institutions invested in digital innovation and economic development. We show that access to LLMs leads to a more productive and advanced OSS ecosystem, which, in many domains, drives the creation of cutting-edge technological solutions with no commercial equivalent (Blind and Schubert 2024). Considering the increasingly impactful nature of OSS on modern economies (Blind and Schubert 2024; Ghosh 2007; Lerner and Tirole 2002), it is essential that policymakers view LLM accessibility as a strategic enabler of innovation. Facilitating equitable and reliable access to LLMs—particularly for contributors to OSS projects—can help foster local talent development, accelerate digital transformation, and stimulate inclusive economic growth.

---

[33] An LLM interruption refers to a temporary loss of access to large language model services—such as those from OpenAI or Anthropic—due to outages or API failures that prevent users from using these tools in their workflows. Despite ongoing advancements in LLM technology, such interruptions continue to occur—for example: https://techinformed.com/openai-global-outage-chatgpt-api-sora/

[34] Many organizations rely on external LLM providers (https://lsvp.com/stories/remarkably-rapid-rollout-of-foundational-ai-models-at-the-enterprise-level-a-survey/)

# Appendices

## Tables

### Table A1. Project Initiation of Software Developers

| Variables | Project Initiation |
|---|---|
| $Treatment \times AfterBan$ | 1.021 |
|  | (0.029) |
| $Treatment \times AfterLift$ | 1.099*** |
|  | (0.032) |
| Working Days | 1.045** |
|  | (0.022) |
| Constant | 0.555*** |
|  | (0.056) |
| Observations | 260,399 |
| User FE | YES |
| Week FE | YES |
| Pseudo $R^2$ | 0.238 |

**Note:** The dependent variable is Project Initiation. On GitHub, this activity is recorded as *Creating a Repository*. The DiD coefficients represent the incidence rate ratios (IRRs), with 1 being the baseline. Robust standard errors are clustered at the User ID level and are reported in parentheses. The significance levels are denoted by ***p < 0.01, **p < 0.05, *p < 0.1.

### Table A2. GitHub Terminology

| Term | Definition |
|---|---|
| Repository | A directory or space for storing projects and files, along with each file's revision history. |
| Gist | A tool for sharing small pieces of code, focusing on individual files rather than entire projects. |
| Commit | A change or a set of changes to files within a repository. Each commit is a snapshot of the repository at a specific point in time. |
| Pull | A mechanism that allows contributors to request that their changes be merged into the project. |
| Issue | A feature that enables users to report problems and suggest enhancements related to a project. |
| Discussion | A forum-like space that enables open conversation about various aspects of a project. |

*Table A2: Definitions of GitHub-specific terms*

**Table A3. Summary Statistics of User Profile Data**

| Variable | Count | | | Mean | | | SD | | | Min | | | Max | | |
|---|---|---|---|---|---|---|---|---|---|---|---|---|---|---|---|
| | Italy | France | Portugal | Italy | France | Portugal | Italy | France | Portugal | Italy | France | Portugal | Italy | France | Portugal |
| Company Affiliation (Indicator) | 21151 | 53918 | 12953 | 0.47 | 0.49 | 0.46 | 0.50 | 0.50 | 0.50 | 0 | 0 | 0 | 1 | 1 | 1 |
| Number of Public Repositories | 21151 | 53918 | 12953 | 18.26 | 21.59 | 19.65 | 31.13 | 42.25 | 44.31 | 0 | 0 | 0 | 1278 | 3966 | 2979 |
| Number of Public Gists | 21151 | 53918 | 12953 | 2.12 | 2.50 | 2.26 | 14.11 | 14.55 | 10.34 | 0 | 0 | 0 | 868 | 1818 | 406 |
| Number of Followers | 21151 | 53918 | 12953 | 17.02 | 18.69 | 20.26 | 161.41 | 137.03 | 211.98 | 0 | 0 | 0 | 17929 | 15561 | 18673 |
| Number of Accounts Followed | 21151 | 53918 | 12953 | 14.19 | 13.03 | 27.68 | 129.61 | 153.66 | 1140.85 | 0 | 0 | 0 | 14188 | 28280 | 128333 |
| User Tenure (Months) | 21151 | 53918 | 12953 | 69.28 | 73.50 | 69.30 | 42.97 | 44.37 | 44.87 | 0 | 0 | 0 | 183 | 183 | 183 |

*Table A3: Summary statistics of user profile data in the panel, by country*

# Figures

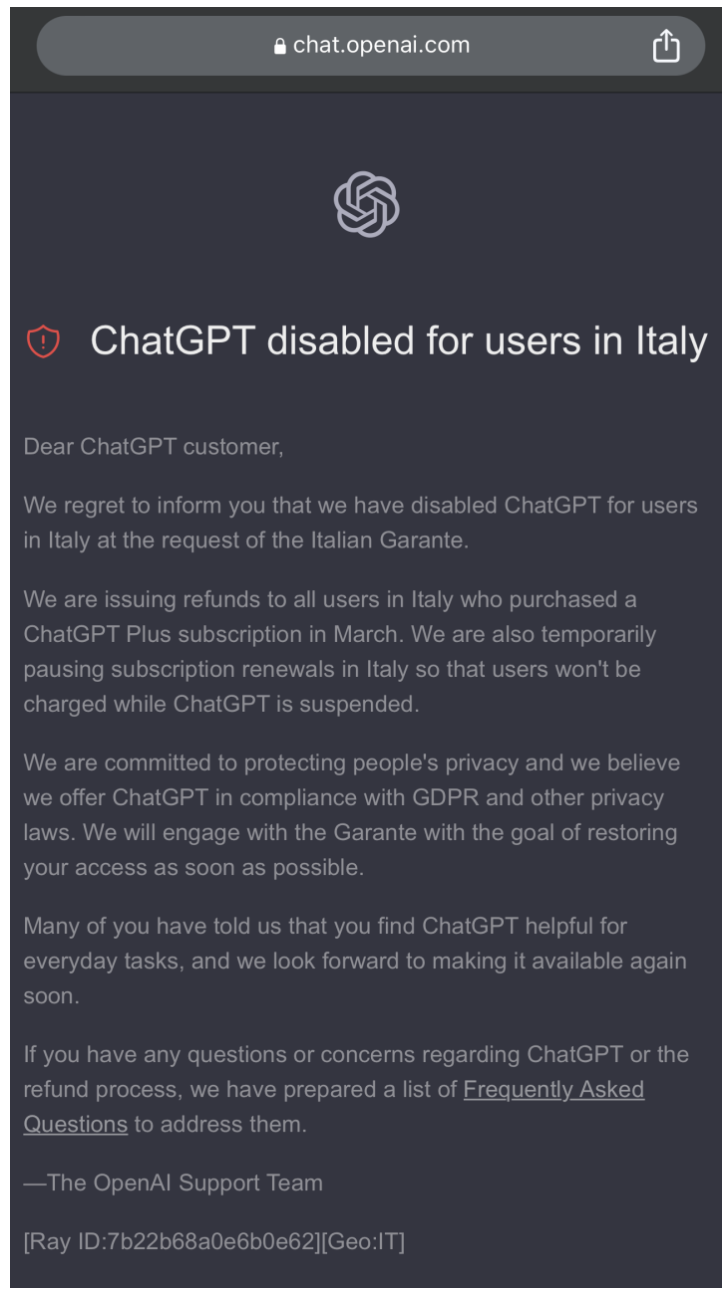

*Figure A1: ChatGPT ban in Italy*

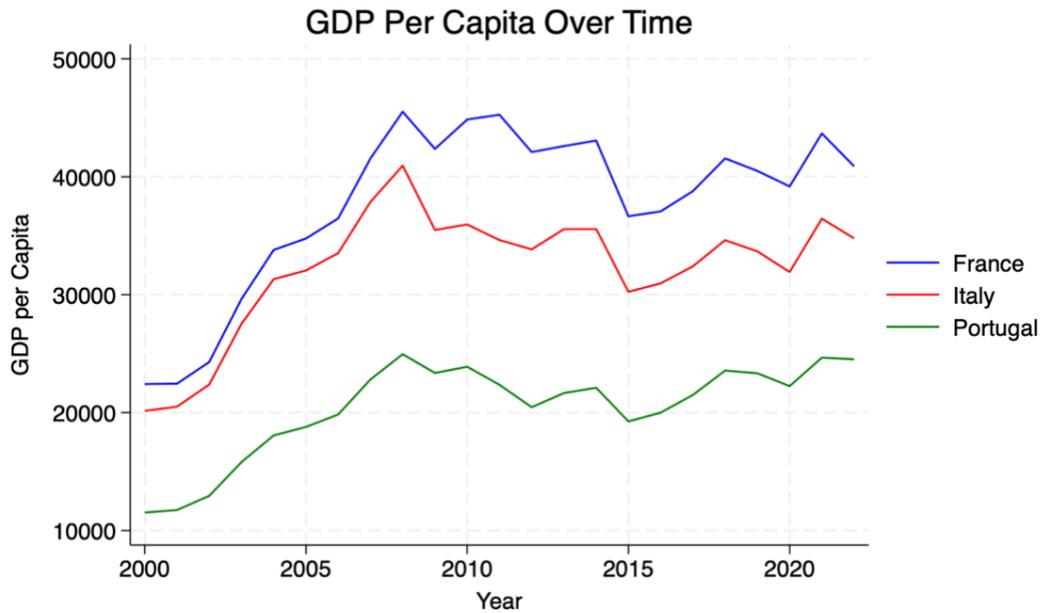

*Figure A2: GDP per capita for Italy, France, and Portugal, from 2000 to 2022*

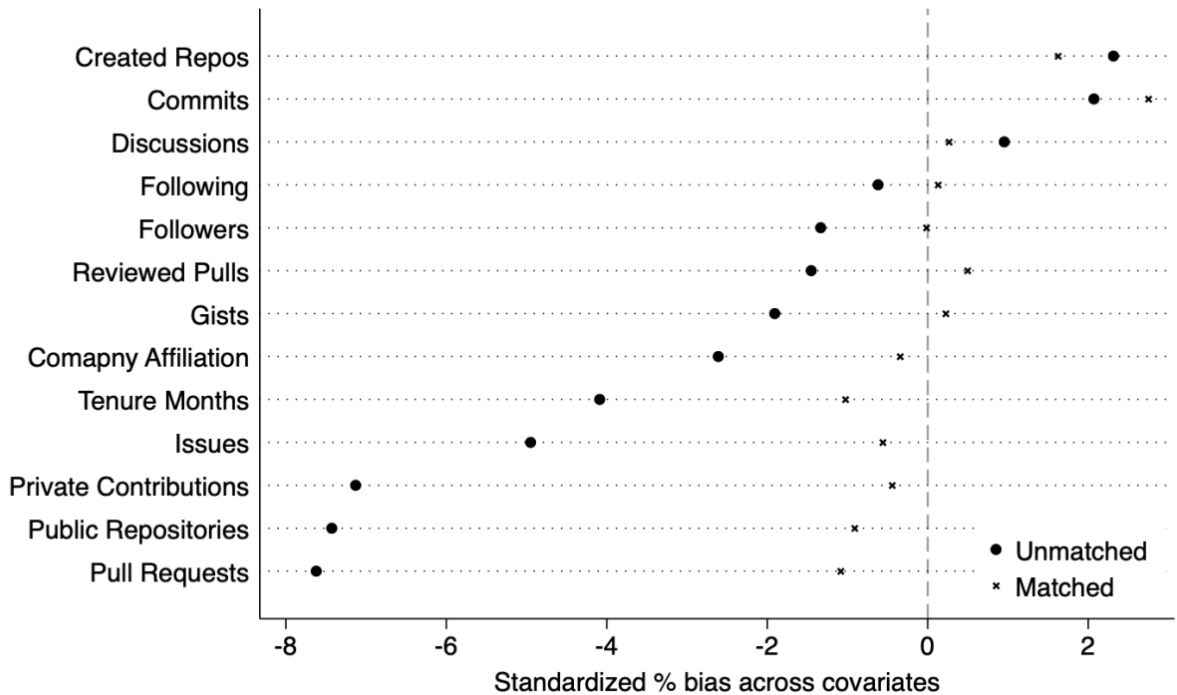

*Figure A3: Percentage reduction in standardized bias for each covariate after matching*

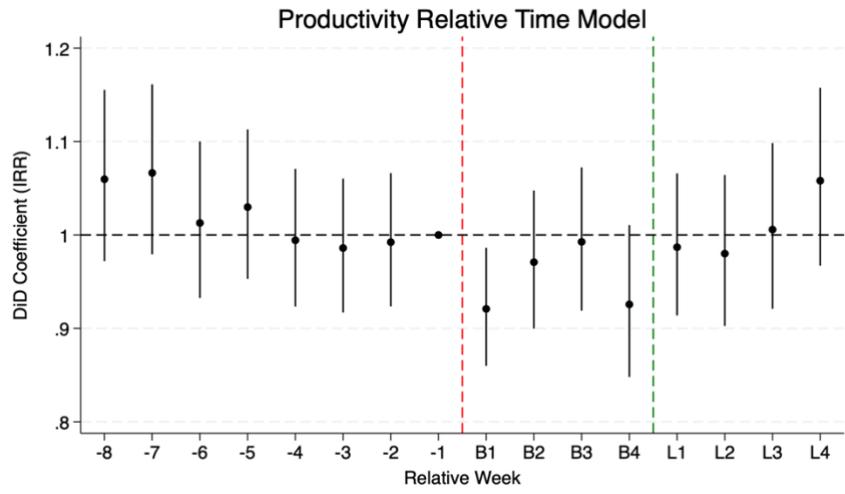

*Figure A4: Effect on the productivity of users in Italy, by week*

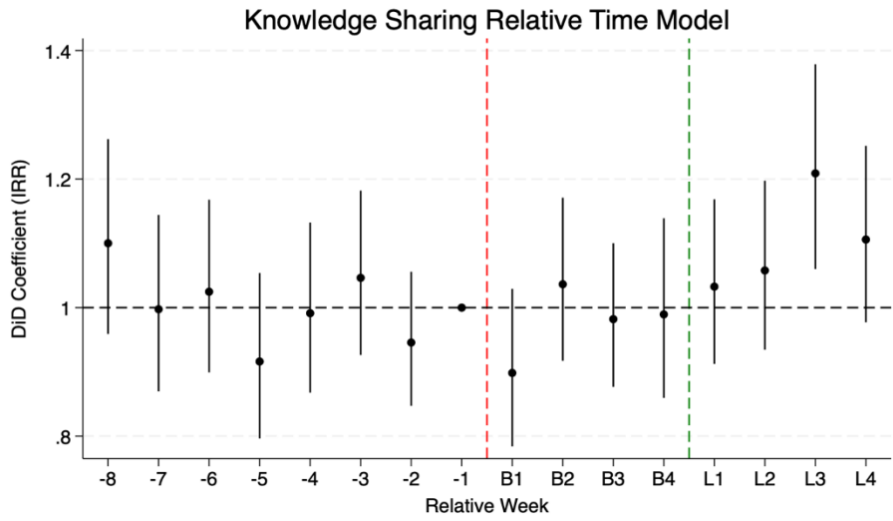

*Figure A5: Effect on the knowledge sharing of users in Italy, by week*

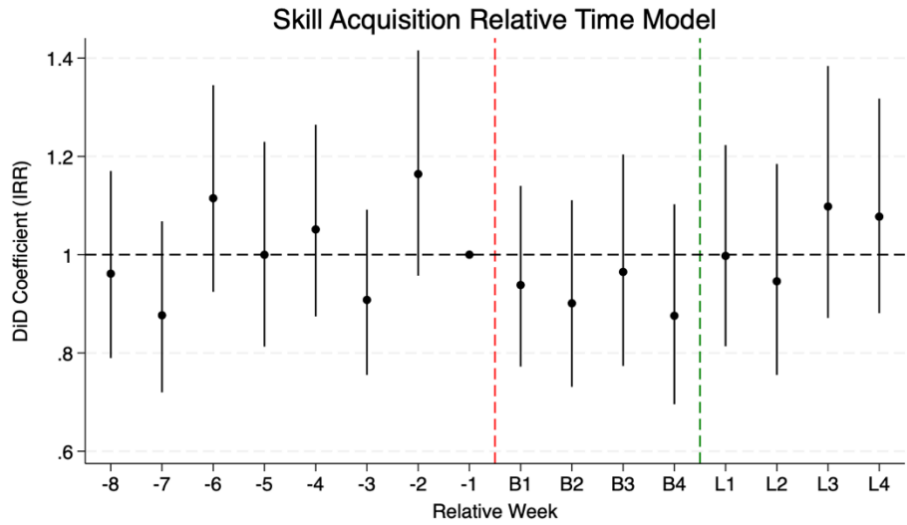

*Figure A6: Effect on the skill acquisition of users in Italy, by week*

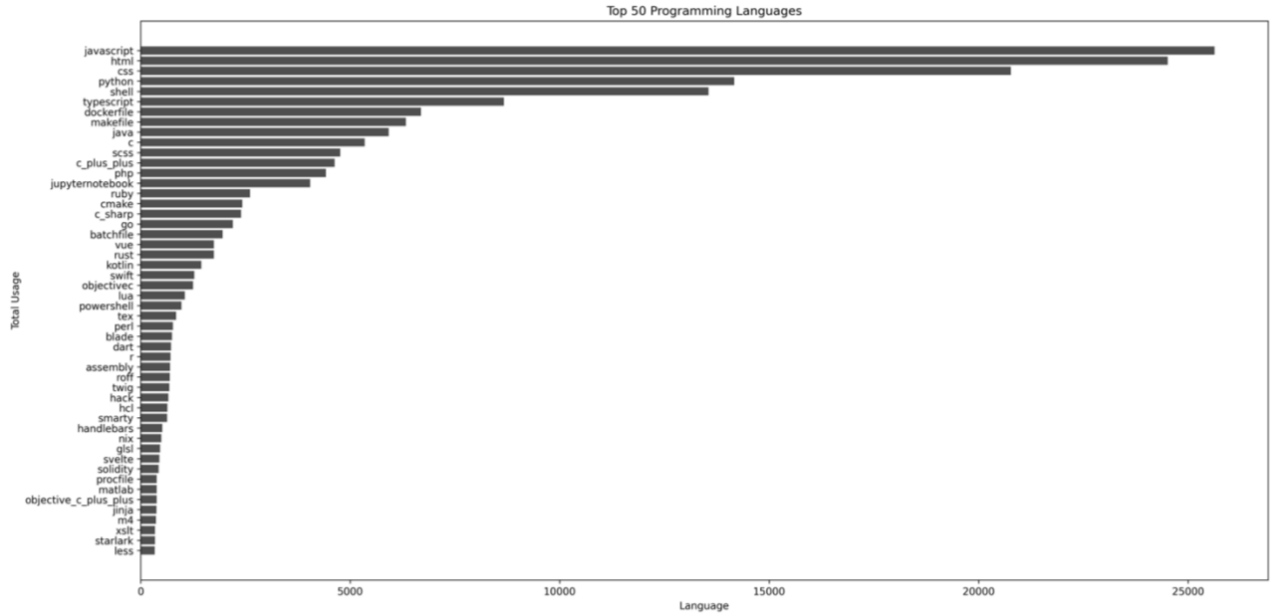

*Figure A7: Top 50 programming languages used by all users in the panel prior to the ban*

## Section A1: Definition of the Skill Acquisition Variable

We derived a new variable called *New Languages* that captures the number of new programming languages a user employed in their different projects during their tenure. Specifically, let the set of all unique languages that user $i$ has used in their first $p-1$ projects during the first $t-1$ time periods be denoted as $L_{i,p-1}^{t-1} = \{l_1, l_2, \ldots, l_n\}$. Suppose user $i$ uses a new set of $m$ languages $\{l_{n+1}, l_{n+2}, \ldots, l_{n+m}\}$ in their new project $p$. Based on these sets, we define *New Languages* as a variable that takes the value $m$ for each project $p$. We refer to the resulting dataset the *Languages Dataset*.

## Section A2: Circumvention Tools Adoption

The imposition of the ChatGPT ban in Italy presented a unique scenario that significantly impacted user behavior, as demonstrated in Section 5 (Main Results). While the ban aimed to restrict access to the tool, it inadvertently prompted a range of adaptive responses from users. In the face of these restrictions, users may have sought methods to circumvent the ban and access ChatGPT, including the use of Virtual Private Networks (VPN). While we cannot user-level adoption of such tools, population-level anecdotal evidence of such efforts (presented in Figure A8) shows a significant increase in searches for the term "VPN" during the ban period, followed by a return to pre-ban levels once the ban was lifted. It is important to note that this increase reflects population-level behavior across all *internet users* in Italy, not just software developers.

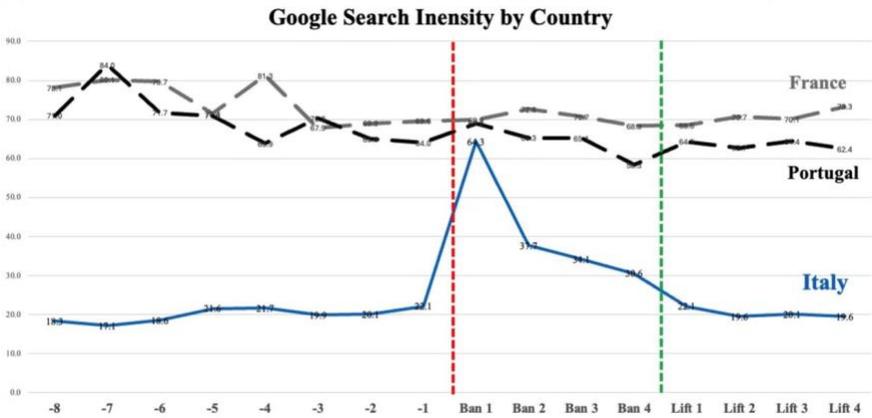

*Figure A8: Google Search Intensity for the term "VPN" for each country*

The potential adoption of circumvention tools is important, as their use allows users in the treated country to continue accessing ChatGPT despite the ban, which can weaken the estimated impact in our Difference-in-Differences (DiD) analysis. If a portion of users in the treated country bypassed the restriction, then not all users were in fact exposed to the treatment as assumed. This means the estimated effect of the ban is averaged over a group that includes both treated and untreated users, which can dilute the measured impact and lead to an underestimation of the true treatment effect.[35]

To minimize the impact of circumvention tool adoption on our analysis, we restrict the observation window to just the first week of the ban. By narrowing the timeframe, we reduce the likelihood that a substantial number of users had sufficient time to adopt these mechanisms, thereby preserving the integrity of our findings.[36] We use the same empirical framework presented in Section 4.2 for estimation.

The results are presented in Table A4. Columns (1), (3), and (5) show outcomes from regressions conducted on data limited to the first week following the ban. Columns (2), (4), and (6) provide results from regressions on the complete dataset, as detailed in Section 5 for easier comparison. The regression analysis of the productivity, knowledge sharing, and skill acquisition variables using the limited-window data yields results that are both qualitatively and quantitatively similar to those obtained from the full dataset. Specifically, as shown in column (1) of Table A4, the regression using the limited-window sample indicates that the ban on ChatGPT resulted in an 8.7% decrease in productivity (calculated as $0.913 - 1$) for users in Italy immediately following the ban. This finding is consistent with the 6.4% decrease observed in the full-data analysis, both in terms of direction and magnitude. Furthermore, similar to the complete-dataset findings, the DiD coefficient for the knowledge sharing variable remains non-significant at the 5% level when using the limited-window data, as shown in column (3) of Table A4. Lastly, the DiD coefficient for skill acquisition is quantitatively similar to the coefficient from the full observation window, albeit insignificant, as shown in Table A4, column (5).

---

[35] Specifically, let $n_j$ denote the total number of users in treated country $j$, all of whom are assumed to belong to the treatment group. Suppose $n_j^c > 0$ users in country $j$ use circumvention tools to access ChatGPT during the ban. In this case, the assumption that all users in the treated country were exposed to the treatment no longer holds, since $n_j > n_j^*$, where $n_j^* = n_j - n_j^c$ represent the true number of treated users (and considering $n_j^c > 0$, $n_j^* \neq n_j$). This discrepancy can lead to an underestimation of the magnitude of the estimated average treatment effect (ATT), as the identified effect is averaged over a larger value ($n_j$) which is the full sample size, rather than the actual number of treated users $n_j^*$, thereby diluting the observed effect.

[36] Assuming $n_{j,t}^c$ represents the number of users in treated country $j$ who adopted circumvention tools during week $t \in \{b_1, b_2, b_3, b_4\}$ of the four-week ban period. the total number of users who adopted such tools during the entire ban period is given by $n_j^c = n_{j,b_1}^c + n_{j,b_2}^c + n_{j,b_3}^c + n_{j,b_4}^c$. By limiting the post-ban observation window to just the first week, we effectively set $n_j^c = n_{j,b_1}^c \leq n_{j,b_1}^c + n_{j,b_2}^c + n_{j,b_3}^c + n_{j,b_4}^c$, thereby minimizing potential contamination from circumvention tool adoption in later weeks.

**Table A4. Regression Results for the Immediate Effects of the Ban**

|  | Productivity | | Knowledge Sharing | | Skill Acquisition | |
| --- | --- | --- | --- | --- | --- | --- |
| Sample Construction | (1) Immediate Effect | (2) Overall Effect | (3) Immediate Effect | (4) Overall Effect | (5) Immediate Effect | (6) Overall Effect |
| $Treatment \times AfterBan$ | 0.913*** (0.027) | 0.936*** (0.021) | 0.880* (0.059) | 0.972 (0.035) | 0.937 (0.088) | 0.916* (0.047) |
| $Treatment \times AfterLift$ | – | 0.993 (0.026) | – | 1.096** (0.050) | – | 1.013 (0.054) |
| Working Days | 1.037 (0.044) | 1.069*** (0.018) | 1.214** (0.113) | 1.073*** (0.027) | 1.012 (0.115) | 0.999 (0.045) |
| Constant | 12.727*** (2.677) | 10.375*** (0.853) | 1.669 (0.776) | 2.787*** (0.336) | 0.100*** (0.063) | 0.092*** (0.028) |
| Observations | 171,628 | 298,632 | 66,066 | 132,598 | 39,991 | 70,766 |
| User FE | YES | YES | YES | YES | YES | YES |
| Week FE | YES | YES | YES | YES | YES | YES |
| Pseudo $R^2$ | 0.543 | 0.507 | 0.541 | 0.535 | 0.474 | 0.479 |

**Note:** The dependent variable is productivity in columns (1) and (2), knowledge sharing in columns (3) and (4), and skill acquisition in columns (5) and (6). The DiD coefficients represent the incidence rate ratios (IRRs), with 1 being the baseline. Robust standard errors are clustered by User ID and are reported in parentheses. The significance levels are denoted by ***p < 0.01, **p < 0.05, *p < 0.1.

Based on the analysis presented in this section, we conclude that despite the potential adoption of circumvention tools by some users within the treated country, the overall impact of the ChatGPT ban and its subsequent lifting on productivity, knowledge sharing, and skill acquisition activities of users within the treated country remains well-identified within the model proposed in Section 4.2. Accordingly, the results presented in Section 5 remain robust.

**Section A3: Programming Languages Clusters**

**1) General-Purpose:** General-purpose languages are versatile and designed to address a wide array of programming tasks, including application development, software engineering, and automation. These languages are not confined to specific domains but can be applied across multiple areas, from desktop applications to back-end (server-side logic) services (Bhatt and Pahade 2021; Fayed et al. 2020). These programming languages include Java, Python, C#, Go, Perl, and Kotlin (Raschka et al. 2020).

**2) Web Development:** Web development languages are mainly designed for the creation and management of web-based applications, including both front-end (user interface) and back-end (server-side logic) development (Jain et al. 2024). These languages are tailored to handle the unique requirements of web technologies, such as HTTP communication, browser rendering, and interactive content. This specialization distinguishes them from other categories like general-purpose or system programming, which are not inherently web-focused. These programming languages and technologies include JavaScript, HTML, CSS, TypeScript, SCSS, PHP, Ruby, Vue.js, Blade, Dart, Twig, Smarty, Handlebars.js, Less.js, Svelte, and Dart.

**3) System Programming:** System programming languages are specialized for low-level operations that directly interact with hardware and manage system resources. They are commonly used in developing operating systems, device drivers, and embedded systems, where performance and efficiency are critical. Unlike general-purpose or web development languages, system programming languages focus on precise control over hardware rather than application logic or web functionality. These programming languages include Assembly, C, C++, and Rust (Bugden and Alahmar 2022).

**4) Scientific Computing:** Scientific computing languages are optimized for tasks requiring high-performance numerical computation, such as linear algebra operations on multidimensional arrays (Raschka et al. 2020), data analysis, and complex mathematical modeling. These languages are widely adopted in fields that demand precision and speed in scientific calculations, such as engineering, physics, and data science. Their optimization for computational performance makes them distinct from other categories like web development or system programming. These programming languages include MATLAB and R (Perkel 2018; Wang et al. 2022; Zúñiga-López et al. 2020).

**5) DevOps and Configuration:** Languages in this category are specifically designed for automating the provisioning, configuration, and orchestration of software infrastructure. These languages and tools enable developers and operations teams to define and manage infrastructure as code (IaC), automate complex deployment pipelines, and ensure consistency across different environments. Unlike general-purpose and system programming languages, which are intended for application development or low-level hardware interaction, DevOps and Configuration languages and technologies focus on operational automation and system management. These languages and technologies include DockerFile, Makefile, CMake, Batchfile, Nix, Procfile, PowerShell, and Starlark.

**6) Templating and Markup:** Templating and markup languages are specifically designed to define the structure and presentation of documents and web content. These languages and technologies allow the separation of content from design, which enables the dynamic generation of content without altering the underlying structural templates. This exclusive focus on document formatting and content presentation differentiates them from other categories that handle more functional or computational tasks, such as system programming or scientific computing. These languages and technologies include TeX, Roff, XSLT, and Jinja.

**7) Domain-Specific:** Domain-specific languages (DSLs) are designed for specialized tasks or industries, such as telecommunications, financial modeling, or game development. These languages are purpose-built to solve narrowly defined problems within specific domains. This specialization makes them distinct from other categories such as general-purpose or scientific computing category languages, which are used across broader applications. These languages and technologies include Swift, Objective-C, Objective-C++, Solidity, Hack, Lua, HCL, m4, and GLSL.

## Section A4: Full-Sample Language Cluster Learning Results

Following the approach in Section 8, we repeated the analysis using the full sample, which includes users across all experience levels. The results, shown in Table A5, indicate that neither the ban nor its lift had a significant effect on learning any language cluster, except for the domain-specific cluster. Potential reasons for this reduction are explained in Section 8.

**Table A5. Regression Results for New Programming Language Acquisition by Cluster**

|  | (1) General Purpose | (2) Web Development | (3) System Programming | (4) Scientific Computing | (5) DevOps and Configuration | (6) Templating and Markup | (7) Domain-Specific |
|---|---|---|---|---|---|---|---|
| $Treatment \times AfterBan$ | 1.187 (0.144) | 0.879 (0.081) | 0.882 (0.154) | 1.234 (0.573) | 0.908 (0.116) | 1.026 (0.281) | 0.702* (0.150) |
| $Treatment \times AfterLift$ | 0.906 (0.121) | 1.041 (0.107) | 0.794 (0.149) | 0.709 (0.316) | 0.889 (0.121) | 1.013 (0.298) | 1.044 (0.203) |
| Obs. | 23,092 | 41,213 | 13,235 | 2,815 | 22,933 | 7,298 | 15,014 |
| User FE | YES | YES | YES | YES | YES | YES | YES |
| Week FE | YES | YES | YES | YES | YES | YES | YES |
| Pseudo $R^2$ | 0.22 | 0.35 | 0.29 | 0.20 | 0.33 | 0.37 | 0.39 |

**Note:** In column (1), the dependent variable represents the total number of new programming languages learned within the General-Purpose category as the dependent variable. Columns (2) through (7) represent the same for Web Development, System Programming, Scientific Computing, DevOps and Configuration, Templating and Markup, and Domain-Specific categories, respectively. Control terms are omitted for brevity. The DiD coefficients represent the incidence rate ratios (IRRs), with 1 being the baseline. Robust standard errors are clustered at the User ID level and are reported in parentheses. The significance levels are denoted by ***p < 0.01, **p < 0.05, *p < 0.1.